\documentclass[12pt]{elsarticle}                                          
\usepackage{axodraw4j}
\usepackage{hyperref}
\usepackage{graphicx}                                                         
\usepackage{a41}                                                                
\usepackage{xcolor}                     
\usepackage{slashed}
\usepackage[rflt]{floatflt}
\usepackage{float}
\usepackage{hyperref}
\hypersetup{
colorlinks=true,
linkcolor=blue,
filecolor=blue,
urlcolor=mygreen
}
\usepackage{breakurl} 
\usepackage{times}
\usepackage{array}
\usepackage[english]{babel}
\usepackage[latin1]{inputenc}
\usepackage[T1]{fontenc}
\usepackage{ae}
\usepackage{url}
\usepackage{amsmath, amsthm, amssymb}
\usepackage{slashed}

\usepackage{rotating}
\usepackage{graphicx}
\usepackage{comment}
\newcounter{mmacnt}
\def\restartmma{\setcounter{mmacnt}{0}}
\restartmma \catcode`|=\active
\def|#1|{\mathrm{#1}}
\catcode`|=12
\newenvironment{mma}{
\par\smallskip
\catcode`|=\active
\parskip=0pt\parindent=0pt 
\small
\def\In##1\\{%
\def\linebreak{\hfill\break\null\qquad}%
\refstepcounter{mmacnt}
\hangindent=2.5em\hangafter=0
\leavevmode
\llap{\tiny\sffamily In[\arabic{mmacnt}]:=\kern.5em}%
\mathversion{bold}\footnotesiVe$
\displaystyle##1$\normalsiVe
\mathversion{normal}\par
 }%
\def\Print##1\\{%
\def\linebreak{\hfill\break}%
\hangindent=2.5em\hangafter=0
\leavevmode ##1\par}%
\def\Out##1\\{%
\def\linebreak{$\hfill\break\null\hfill$}%
\kern\abovedisplayskip\par
\hangindent=2.5em\hangafter=0
\leavevmode
\llap{\tiny\sffamily Out[\arabic{mmacnt}]=\kern.5em}
\footnotesiVe$\displaystyle##1$
\normalsiVe\hfill\null\par
\kern\belowdisplayskip
}%
\def\Warning##1##2\\{%
\def\linebreak{\hfill\break}%
\hangindent=2.5em\hangafter=0
\leavevmode
{\scriptsiVe##1 : ##2}\par}%
}{%
\par\smallskip
}

\usepackage{color}
\newenvironment{fshaded}{%
\MakeFramed {\FrameRestore}
}%
{\endMakeFramed}

\makeatletter
\def\ps@pprintTitle{%
\let\@oddhead\@empty
\let\@evenhead\@empty
\def\@oddfoot{\reset@font\hfil\thepage\hfil}
\let\@evenfoot\@oddfoot
}
\makeatother
\usepackage{tcolorbox}
\usepackage{slashed}
\begin{document}  
\begin{frontmatter}
\title{\textbf{One-loop induced
contributions to the rare
decay of $A_0 \rightarrow h_0h_0\gamma$
in Two Higgs Doublet Models}}
\author[1,2]{Dzung Tri Tran}
\author[3,4]{L. T. Hue}
\author[5,6]{Thanh Huy Nguyen}
\author[5,6]{Vo Quoc Phong}
\author[1,2]{Khiem Hong Phan}
\ead{phanhongkhiem@duytan.edu.vn}
\address[1]{\it Institute of Fundamental
and Applied Sciences, Duy Tan University,
Ho Chi Minh City $70000$, Vietnam}
\address[2]{Faculty of Natural Sciences,
Duy Tan University, Da Nang City $50000$,
Vietnam}
\address[3]{\it Subatomic Physics Research Group,
Science and Technology Advanced Institute,
Van Lang University, Ho Chi Minh City, Vietnam}
\address[4]{\it
Faculty of Applied Technology, School of Technology,
Van Lang University, Ho Chi Minh City, Vietnam}
\address[5]{\it
Department of Theoretical Physics, 
University of Science, 
Ho Chi Minh City $700000$, Vietnam}
\address[6]{\it Vietnam National University, 
Ho Chi Minh City $700000$, Vietnam}
\pagestyle{myheadings}
\markright{}
\allowdisplaybreaks
\begin{abstract}
The analytic expressions for one-loop contributions to the rare decay process $A_0 \rightarrow h_0h_0\gamma$ within the CP-conserving of  Two Higgs Doublet Models are first reported in this paper.  Analytic results are presented in term of scalar one-loop Passarino-Veltman functions following the standard output of the packages~{\tt LoopTools}  and {\tt Collier}. In this context, physical results for the computed process are easily generated by using one of 
these  packages. The numerical checks are proposed to verify for the analytic results in this paper. The checks rely on the renormalization conditions that the decay amplitude must be the ultraviolet finiteness and  infrared finiteness. The amplitude consisting of an external photon always obeys the Ward identity. This will be confirmed numerically in this article. In phenomenological  results, the decay rates of $A_0 \rightarrow h_0h_0\gamma$ are evaluated at several points in the allowed regions of the  parameter space. Furthermore, the  differential decay widths with respect to the invariant mass of Higgs-pair in final states are studied. 
\end{abstract}
\end{frontmatter}
\section{Introduction}
After discovering the  Standard-Model-like Higgs boson (SM-like Higgs) $h_0$ at the 
Large Hadron Collider (LHC)~\cite{ATLAS:2012yve,CMS:2012qbp}, the standard model (SM)  theory of particle  physics is being well-established.  In spite of the great SM's success in describing most of  the experimental data, the structure  of the scalar Higgs sector in the SM  is still an unknown question. There is no theoretical principle  for the minimum of  scalar Higgs sector selection in the SM. In many physics beyond the SM (BSM), the  Higgs potential is extended by adding new scalar singlets and/or multiplets. Subsequently,  many additional scalar bosons appear, such as neutral CP-even, CP-odd,  and charged Higgs bosons. In the context of experimental searches, the precise measurements for the decay widths and the production cross-sections for all the above-mentioned scalar particles could play important roles in exploring the Higgs sector, understanding deeply the origin of the electroweak symmetry breaking mechanism (EWSB) as well as probing new physics signals. Among the mentioned scalar particles, decay and processes of CP-odd ($A_0$) are of considerable interest. Recently, direct production of  a light CP-odd Higgs boson  has been performed at the Tevatron and LHC~\cite{Dermisek:2009fd}. Searches  for the decay $A_0 \rightarrow Zh_0$ in $pp$ collisions have been carried out at the LHC~\cite{ATLAS:2015kpj}. Probing for a light pseudoscalar Higgs boson in $\mu\mu\tau\tau$ events at the LHC in Refs.~\cite{CMS:2020ffa,ATLAS:2024rzd} and in the di-muon decay channels in $pp$ collisions at $\sqrt{s}=7$ Tev has been reported in Ref.~\cite{CMS:2012fgd}.

In the aspects of theoretical studies, the detailed computations  for one-loop and higher-loop corrections to the decay width of CP-odd Higgs boson are necessary  for matching the higher-precision data at future colliders. One-loop corrections to two-body decay of CP-odd Higgs 
within the CP-conservating of the Two Higgs Doublet Models (THDM) have been performed in Refs.~\cite{Aiko:2022gmz, Kanemura:2017gbi, Kanemura:2019slf,Aiko:2023xui}. One-loop analyses for the most important decay channel $A_0 \rightarrow Zh_0$ have been studied in Refs.~\cite{Akeroyd:2024tbp, Akeroyd:2023kek}. Furthermore, one-loop electroweak corrections to the decay of $A_0$ into a pair of scalar fermion have been calculated in Refs.~\cite{Weber:2003tw,Weber:2003eg} within the context of minimal supersymmetric extension of the Standard Models (MSSM). The CP-odd Higgs  decay rates to two gluons have been evaluated by applying the Pade improvement method in Ref.~\cite{Chishtie:1999dd}. Morerecently, one-loop contributions to $A_0 \rightarrow \ell \bar{\ell}V$ with $V=\gamma,~Z$ in the context of Higgs Extensions of the Standard Models like the THDM and Triple Higgs Models (THM), etc., have been computed in Ref.~\cite{Phan:2024zus}. The CP-odd Higgs boson productions associated with a neutral $Z$ boson at the LHC within the MSSM  framework have beenevaluated in Ref.~\cite{Yin:2002sq}. Evaluating for the productions of CP-odd Higgs boson at future $e^-\gamma$ colliders has been considered in Ref.~\cite{Sasaki:2017fvk}. Moreover, one-loop electroweak corrections  to the process  $e^+ e^- \rightarrow \nu \bar{\nu} A_0$ in the THDM  have computed in Ref.~\cite{Farris:2003pn}.  The CP-odd $A_0$ production at $e^+ e^-$  colliders in  the MSSM with CP-violating phases have Hue{been} computed in Ref.~\cite{Arhrib:2002ti}. Additionally, the CP-odd Higgs  boson production in association with  a neutral  gauge boson $Z$ in high-energy $e^+e^-$ collisions  at one-loop level analyses has been reported in the frameworks of the THDM in Ref.~\cite{Akeroyd:1999gu} and supersymmetric models in Ref.~\cite{Akeroyd:2001aka}.

In this paper, we present the first analytic expressions for one-loop contributions  to the rare decay process $A_0 \rightarrow h_0h_0\gamma$ within the CP-conserving of the THDM. Analytic results are written in terms of scalar one-loop  Passarino-Veltman functions (PV-functions) 
in the standard output of both packages~{\tt LoopTools} and~{\tt Collier}. Numerical checks for the validation of our calculations such as the ultraviolet and infrared finiteness, and the Ward identity of one-loop amplitude are also performed. In phenomenological results,  the decay rates for $A_0 \rightarrow h_0h_0\gamma$ and the differential decay widths with respect to the invariant mass of Higgs-pair in final states  are computed at several points of the allowed regions of the parameter space of the THDM. The paper is arranged as follows. In section 2, we briefly review the THDM. In section 3, the detailed evaluations for one-loop contributions to the decay channel $A_0\rightarrow h_0h_0\gamma$ are presented. Phenomenological results for the calculations are shown in section 4. Important conclusions and outlook of this work are shown in section 5.
\section{Two Higgs Doublet Models}
We first review shortly the model under consideration in the calculation. For a  complete review of this model and relevant phenomenological studies, we refer to Ref.~\cite{Branco:2011iw}. In the structure of the THDM, fermion and gauge sectors are kept the same as those in the SM. With the above extension, the scalar potential, reflecting two gauge and Lorentz invariances, is written in a general form as follows
\begin{eqnarray}
\label{V2HDM}
\mathcal{V}_{\textrm{THDM}}
(\Phi_1,\Phi_2) 
&=&
m_{11}^2\Phi_1^\dagger \Phi_1
+ m_{22}^2\Phi_2^\dagger \Phi_2-
\Big[
m_{12}^2\Phi_1^\dagger \Phi_2
+{\rm H.c.}
\Big]
+ \frac{\lambda_1}{2}
(\Phi_1^\dagger \Phi_1)^2
+\frac{\lambda_2}{2}
(\Phi_2^\dagger \Phi_2)^2
\nonumber
\\
&&
+ \lambda_3(\Phi_1^\dagger \Phi_1)
(\Phi_2^\dagger \Phi_2)
+ \lambda_4(\Phi_1^\dagger \Phi_2)
(\Phi_2^\dagger \Phi_1)
+ \frac{1}{2}
[
\lambda_5~(\Phi_1^\dagger \Phi_2)^2
+ ~{\rm H.c.}
].
\end{eqnarray}
Assuming the CP-conservating
version of the THDM in this work,
all bare parameters in the 
scalar potential are considered to 
be real. Additionally, 
the $Z_2$-symmetry, e.g. $\Phi_1
\leftrightarrow \Phi_1$ and
$\Phi_2\leftrightarrow -\Phi_2$
is implied for the above 
scalar potential up to the 
allowed soft-breaking term  given explicitly as
$m_{12}^2\Phi_1^\dagger
\Phi_2 +{\rm H.c.}$. The parameter
$m_{12}^2$ plays a key 
role of the breaking scale 
for the $Z_2$-symmetry.

For the EWSB, two scalar doublets
are expanded around their VEVs
as follows:
\begin{eqnarray}
\label{representa-htm}
\Phi_k &=&
\begin{bmatrix}
\phi_k^+ \\
(v_k
+
\phi^0_k
+
i
\psi^0_k)
/\sqrt{2}
\end{bmatrix}
\quad \textrm{for}
\quad k=1,2.
\end{eqnarray}
The combined VEV is defined as $v=\sqrt{v_1^2+v_2^2}$, which is  fixed at 
$v \sim 246$ GeV in agreement with the SM limit. The physical particles in the THDM, after the EWSB, include two CP-even Higgs bosons, in which one of them is $h_0$, being the SM-like Higgs boson found at the LHC,  and another one is CP-even Higgs $H$. Furthermore, one also has a CP-odd Higgs ($A_0$) boson and a pair of charged Higgs bosons ($H^\pm$). The masses of all additional scalar bosons can be obtained subsequently by  diagonalizing mass  matrices in their  favor bases. The rotations are  shown in concrete as follows:
\begin{eqnarray}
\begin{pmatrix}
\phi^0_1\\
\phi^0_2
\end{pmatrix}
&=&
\begin{pmatrix}
c_{\alpha}
&
-s_{\alpha}
\\
s_{\alpha}
&
c_{\alpha}
\end{pmatrix}
\begin{pmatrix}
H\\
h_0
\end{pmatrix},
\\
\begin{pmatrix}
\phi_1^{\pm}\\
\phi_2^{\pm}
\end{pmatrix}
&=&
\begin{pmatrix}
c_{\beta}
&
-s_{\beta}
\\
s_{\beta}
&
c_{\beta}
\end{pmatrix}
\begin{pmatrix}
G^{\pm}\\
H^{\pm}
\end{pmatrix},
\\
\begin{pmatrix}
\psi^0_1\\
\psi^0_2
\end{pmatrix}
&=&
\begin{pmatrix}
c_{\beta}
&
-s_{\beta}
\\
s_{\beta}
&
c_{\beta}
\end{pmatrix}
\begin{pmatrix}
G^{0}\\
A_0
\end{pmatrix}.
\end{eqnarray}
The mixing angle $\beta$ is given by $t_{\beta} \equiv \tan \beta= v_2/v_1$. In this circumstance, the neutral and singly charged Goldstone bosons, $G^0$ and $G^{\pm}$, give masses for gauge bosons $Z$ and $W^{\pm}$, respectively. The physical masses of the remaining scalar bosons are then written in  terms of the bare parameters as follows:
\begin{eqnarray}
M_{H^{\pm}}^{2}
&=&
M^{2}-
\frac{1}{2}
\lambda_{45}
v^{2},
\\
M_{A_0}^{2}  &=&
M^{2}-\lambda_{5}v^{2},
\\
M_{h_0}^{2} &=& M_{11}^{2}
s_{\beta-\alpha}^{2}
+ M_{22}^{2}c_{\beta-\alpha}^{2}
+M_{12}^{2}s_{2(\beta-\alpha)},
\\
M_{H}^{2} &=&
M_{11}^{2}c_{\beta-\alpha}^{2}
+M_{22}^{2}s_{\beta-\alpha}^{2}
-M_{12}^{2}s_{2(\beta-\alpha)},
\end{eqnarray}
where $M^{2}=m_{12}^{2}/(s_{\beta}c_{\beta})$,
and
\begin{eqnarray}
M_{11}^{2}&=&
(\lambda_{1}c_{\beta}^{4}
+\lambda_{2}s_{\beta}^{4})v^{2}
+\frac{v^{2}}{2}\;
\lambda_{345}
\; s_{2\beta}^{2}, \\
M_{22}^{2}
&=&
M^{2}
+ \frac{v^{2}}{4}
\Big[
\lambda_{12}
-2\lambda_{345}
\Big]
s_{2\beta}^{2}, \\
M_{12}^{2} &=&
M_{21}^{2}  =
-\frac{v^{2}}{2}
\Big[
\lambda_{1}c_{\beta}^{2}
-\lambda_{2}s_{\beta}^{2}
-
\lambda_{345}\; c_{2\beta}
\Big]
s_{2\beta}.
\end{eqnarray}
Here, the abbreviated notations $\lambda_{ij\cdots} = \lambda_{i}
+\lambda_{j}+\cdots$ 
have been used for simplicity.

Lastly, we discuss the  Yukawa sector in the THDM. It is well-known that the discrete $Z_2$-symmetry is proposed in the THDM  for avoiding Tree-level Flavor-Changing Neutral Currents (FCNCs). Based on the $Z_{2}$-symmetry, the Yukawa interactions appearing in different THDMs are divided into four types, labeled as Type-I, -II, -X, and -Y, respectively, as shown in Table~\ref{Z2-assignment}, which precisely lists all the $Z_2$ charge assignments~\cite{Aoki:2009ha}. Accordingly, the Yukawa interactions are parameterized  in the following general form:
\begin{eqnarray}
{\mathcal L}_\text{Y}
&=&-\sum_{f=u,d,\ell} 
\left( \frac{m_f}{\sqrt{2}v}\xi_h^f{\overline
f}fh_0
+\frac{m_f}{\sqrt{2}v}\xi_H^f{\overline
f}fH
-i\frac{m_f}{v}\xi_A^f{\overline f}
\gamma_5fA_0
\right)\nonumber\\
&&
-\left\{\frac{\sqrt2V_{ud}}{v}\overline{u}
\left(m_u\xi_A^u P_L
+m_d\xi_A^d P_R\right)d\,H^+
+\frac{\sqrt2m_\ell\xi_A^\ell}{v}
\overline{\nu_L^{}}\ell_R^{}H^+
+\text{H.c.}\right\},
\end{eqnarray}
where projection operators are 
$P_{L(R)}$ for left (right)-handed 
of fermions. The elements $V_{ud}$ of the Cabibbo-Kobayashi-Maskawa (CKM) matrix explaining the quark mixing also 
appear in the Yukawa Lagrangian.
\begin{table}[H]
\begin{center}
\begin{tabular}{|c|ccccccc|cccccc|}
\hline\hline
Types & $\Phi_{1}$ & $\Phi_{2}$ &$Q_L$
&$L_L$&$u_R$&$d_R$&$e_R$&$\xi^u_A$ 
&$\xi^d_A$&$\xi^{\ell}_A$ 
&$\xi_h^u$
&$\xi_h^d $
&$\xi_h^{\ell}$
\\\hline\hline
I & $+$ & $-$ 
& $+$ & $+$ & $-$ 
& $-$ & $-$ & 
$\cot\beta$ 
& $-\cot\beta$ 
& $\cot\beta$ 
&
$\dfrac{c_\alpha}{
s_\beta}$
&
$
\dfrac{c_\alpha}{
s_\beta}$
&
$
\dfrac{c_\alpha}{
s_\beta}$
\\ \hline
II & $+$ & $-$ & $+$ & $+$
& $-$ & $+$ & $+$ &
$-\cot\beta$
&$-\tan\beta$
&$-\tan\beta$ 
&
$
\dfrac{c_\alpha}{
s_\beta}$
&
$-
\dfrac{s_\alpha}{
c_\beta}$
&
$-
\dfrac{s_\alpha}{
c_\beta}
$
\\ \hline
X & $+$ & $-$ & $+$
& $+$ & $-$ & $-$ & $+$
& 
$-\cot\beta$ 
& $\cot\beta$
& $-\tan\beta$ 
&
$
\dfrac{c_\alpha}
{s_\beta}$
&
$
\dfrac{c_\alpha}{
s_\beta}$
&
$-
\dfrac{s_\alpha}
{c_\beta}$
\\ \hline
Y & $+$ & $-$
& $+$ & $+$ & $-$
& $+$ & $-$ &
$-\cot\beta$
&$-\tan\beta$
&$\cot\beta$ 
&
$
\dfrac{c_\alpha}
{s_\beta}$
&
$
-
\dfrac{s_\alpha}
{c_\beta}$
&
$
\dfrac{c_\alpha}
{s_\beta}$
\\
\hline
\hline
\end{tabular}
\caption{
\label{Z2-assignment}
The $Z_2$ charge assignments and  
$\xi^f_{A(h)}$ ($f=u, d, \ell$) 
factors corresponding to the  
four THDM types.
While the Yukawa couplings of 
CP-even $H$ to fermion pair
($\xi^f_H$)
are obtained by replacing 
$c_\alpha
\rightarrow s_\alpha$
and vice versa in $\xi^f_h$.
}
\end{center}
\end{table}
The parameter space $\mathcal{P}_{\rm THDM}$
for THDM consists of the following 
free parameters:
\begin{eqnarray}
\label{thdmspace}
\mathcal{P}_{\rm THDM}
=
\{
M_{h_0}^2\sim 125. \textrm{GeV},
M_H^2, M_{A_0}^2, M^2_{H^{\pm}},
m_{12}^2, t_{\beta},
s_{\beta-\alpha}
\}.
\end{eqnarray}
To end this section,  
the current constraints on 
the parameter space of the THDM given in Eq.~\eqref{thdmspace} 
are first summarized. For the subject, 
both theoretical and experimental constraints to 
the model are taken into consideration for 
finding the allowed parameter space of the THDM. 
In the theoretical perspectives, 
the requirements of 
the perturbative regime, the
tree-level unitarity of gauge theory
and the vacuum stability conditions 
of the scalar Higgs potential are 
taken into account~\cite{Nie:1998yn,
Kanemura:1999xf, Akeroyd:2000wc,
Ginzburg:2005dt, Kanemura:2015ska}.
From the experimental data,
the EWPT for the THDM has implicated
at the LEP as reported 
in Refs.~\cite{Bian:2016awe, Xie:2018yiv}.
The bounds from indirect searches  
for the masses range 
of scalar particles in the THDM 
have studied
in Ref.~\cite{Kanemura:2011sj,
Eberhardt:2020dat,Wang:2022yhm,
Karan:2023kyj}.
Implicating one-loop induced for
the SM-like Higgs decay channels
like $h\rightarrow \gamma\gamma$
and $h\rightarrow Z\gamma$ in the
context of THDM have examined in
Refs.~\cite{Chiang:2012qz,
Benbrik:2022bol}. Additionally,
the implications of $W$-boson mass
at the CDF-II in Refs.~\cite{Jung:2022prq,
Ahn:2022xax,Cici:2024fsl} and  
the updated constraints for 
muon $g-2$ anomaly in Ref.~\cite{Iguro:2023tbk}.
Lastly, 
the flavor experimental
data as shown in 
Refs.~\cite{Haller:2018nnx,
Connell:2023jqq,Athron:2024rir}
gives a further constraints on  
$t_{\beta}$,~$M_{H^{\pm}}$.
In detail, the results from 
Refs.~\cite{Haller:2018nnx,
Connell:2023jqq,Athron:2024rir}
pointed out that the small values of
$t_{\beta}$ are favoured
for explaining the flavor experimental
data. For this reason, 
the small values for 
$t_{\beta}$ are also considered 
for complementary discussion
in this work. 
By considering 
all the above-constraints, we can take 
the physical parameters for the THDM
in the regions like
$126$ GeV $\leq M_H \leq 1500$ GeV,
$60+$ GeV $\leq M_{A} \leq 1500$ GeV
and $80$ GeV $\leq M_{H^{\pm}}
\leq 1500$ GeV. The $Z_2$-breaking
parameter can be taken 
as $m_{12}^2 =M_H^2
s_\beta c_\beta$. Our phenomenological 
results studied in the next sections
will be examined in the above-mentioned
parameter space. 
\section{Loop-induced decay of       
$A_0 \rightarrow h_0h_0\gamma$ in THDM}%
\begin{figure}[H]
\centering
\includegraphics[width=12cm, height=5.5cm]
{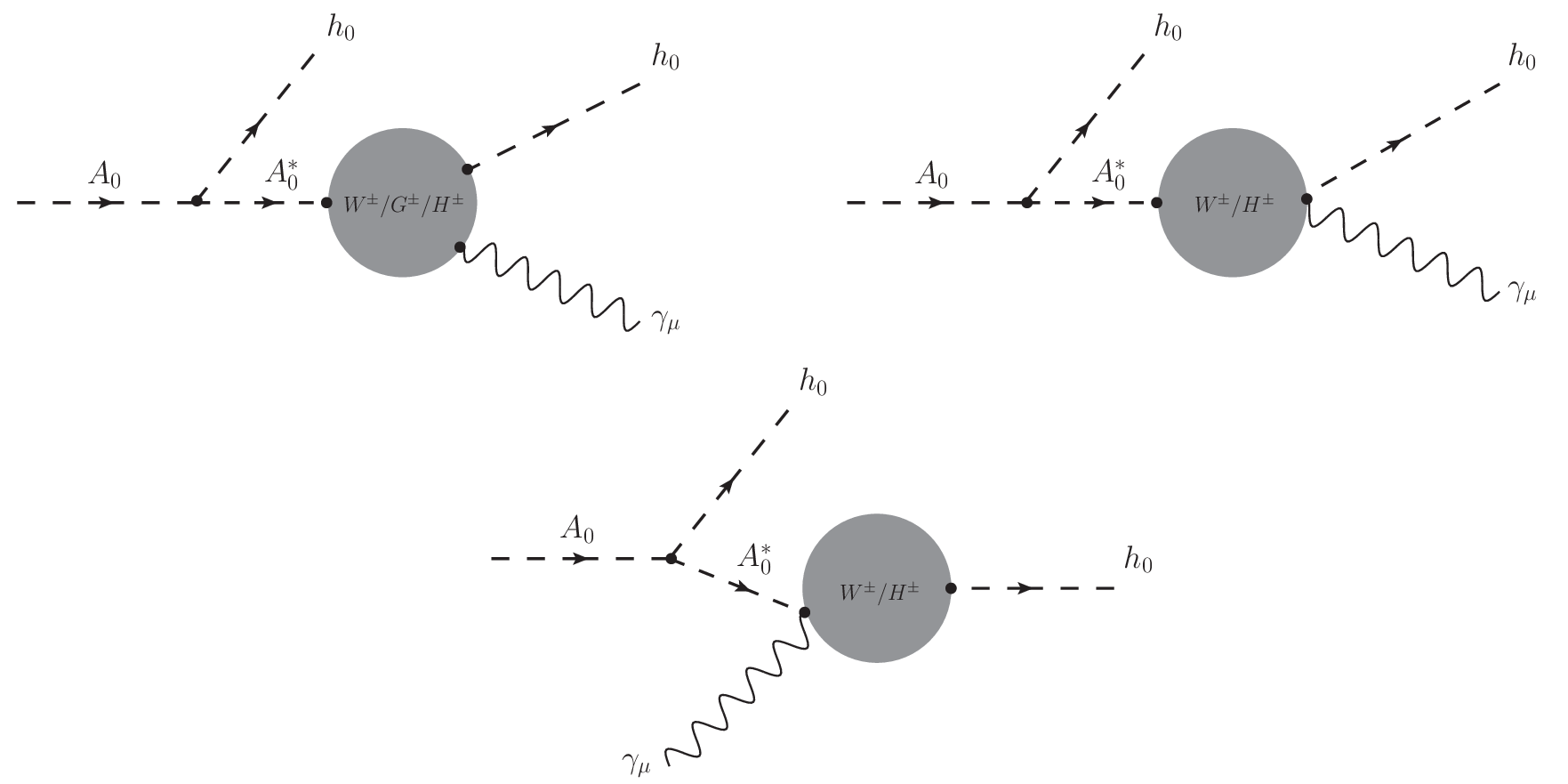}
\\
\includegraphics[width=12cm, height=3cm]
{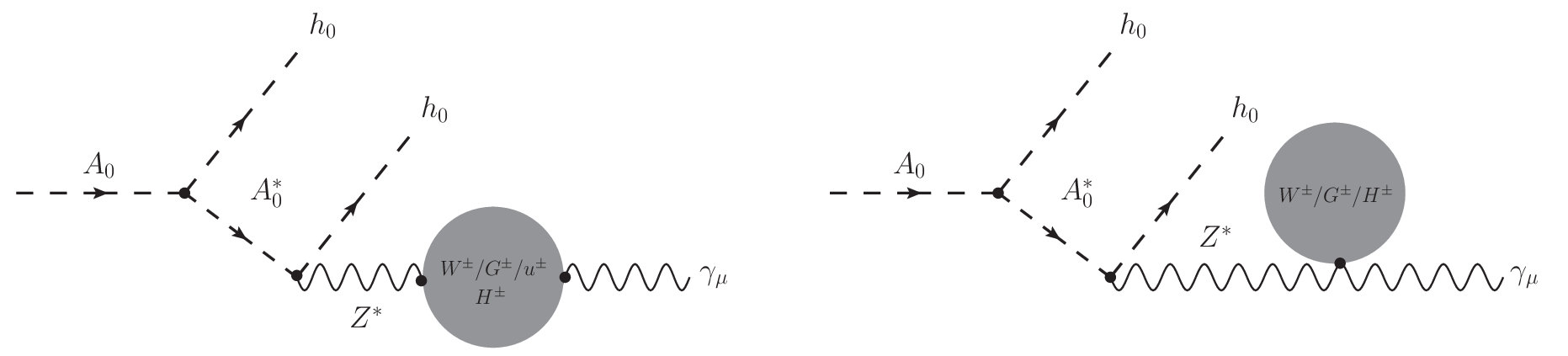}
\caption{
\label{poleA0}
One-loop triangle diagrams including the
mixing of $Z$-$\gamma$ with
$A_0^*$-pole. All particles like 
fermion $f$, vector bosons and scalar
particles are considered to propagate
in the loop diagrams. Within HF gauge,
we have also Goldstone and Ghost particles
exchanging in the loop.}
\end{figure}

\begin{figure}[H]
\centering
\includegraphics[width=12cm, height=5.5cm]
{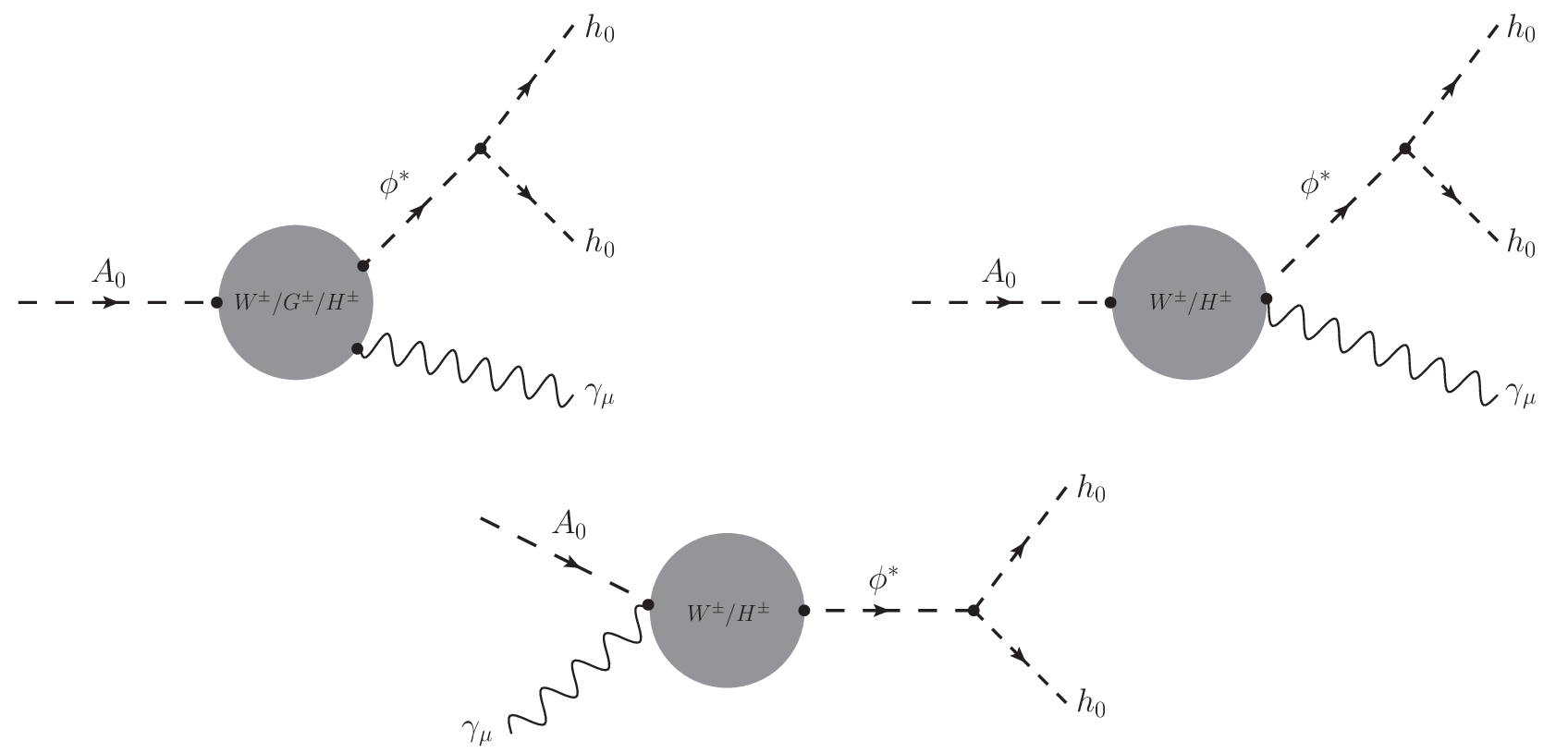}
\\
\hspace*{1.8cm}
\includegraphics[width=12cm, height=3.5cm]
{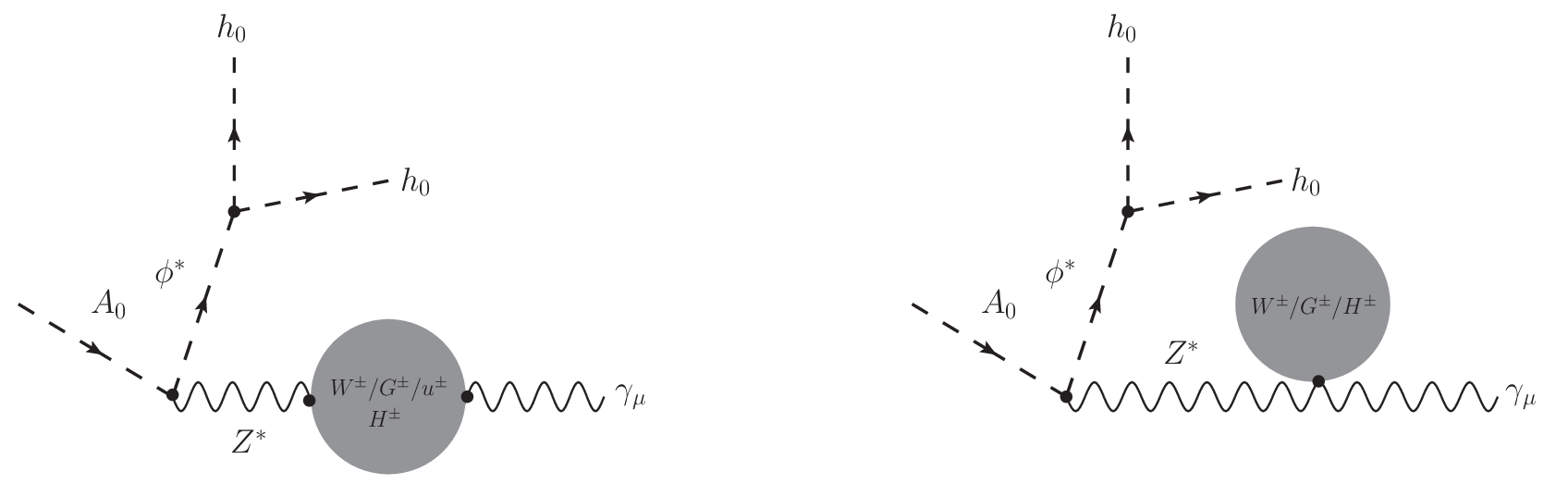}
\caption{
\label{polePHI}
One-loop triangle diagrams
with $\phi$-poles for
$\phi \equiv h_0,~H$, including
the mixing of $Z$-$\gamma$.
All particles like fermion
$f$, vector bosons and scalar
particles are considered to propagate
in the loop diagrams. Within HF gauge,
we have also Goldstone
and Ghost particles
exchanging in the loop.
}
\end{figure}

\begin{figure}[H]
\centering
\includegraphics[width=12cm, height=6cm]
{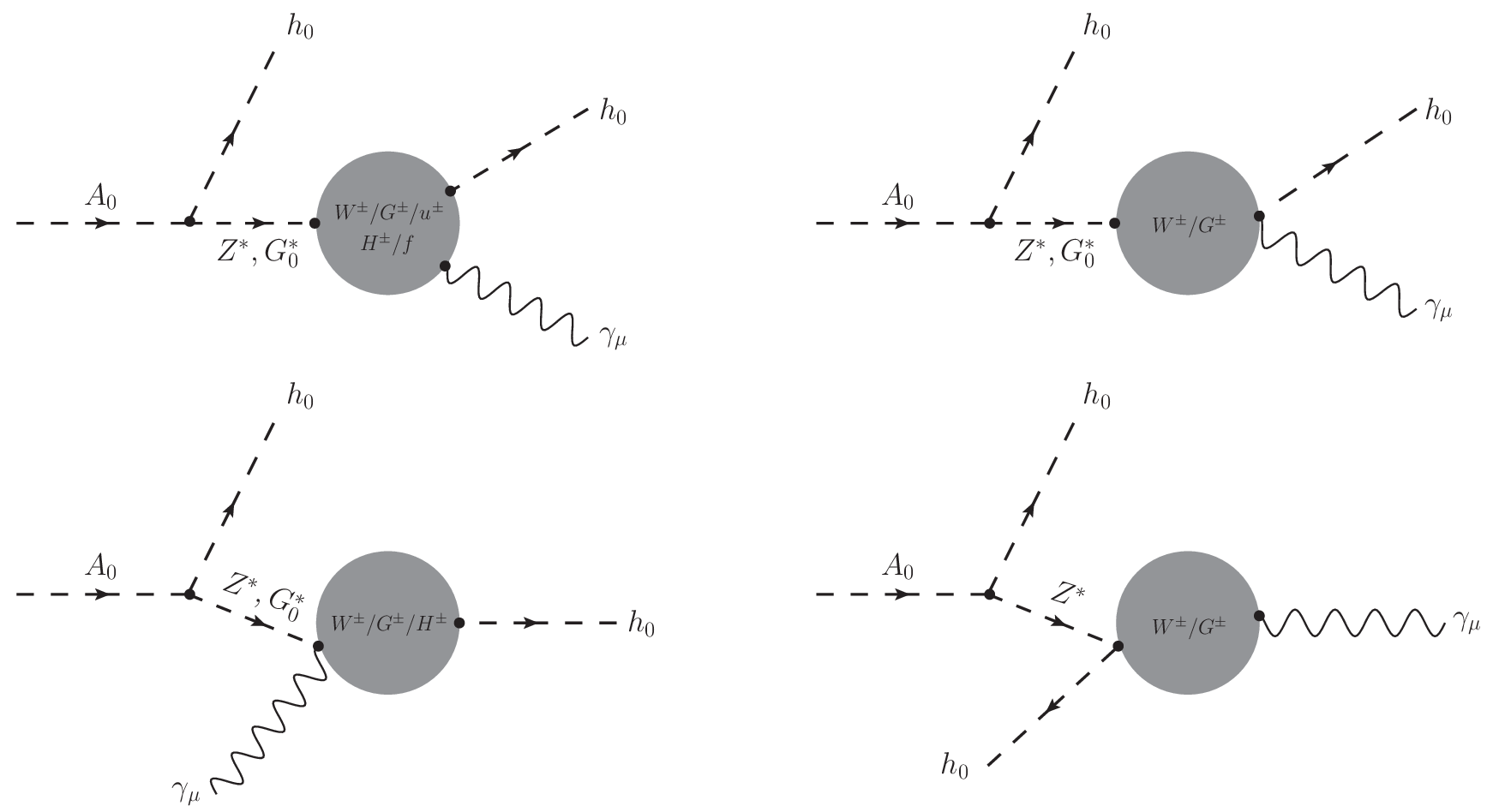}
\\
\includegraphics[width=12cm, height=3cm]
{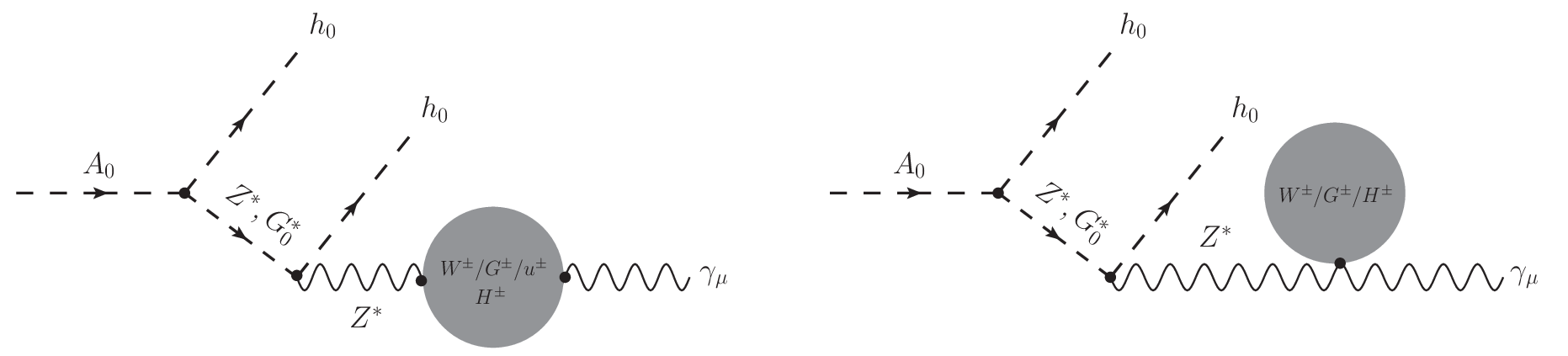}
\caption{
\label{Zpole}
One-loop triangle diagrams
with $Z$-pole including the mixing of
$Z$-$\gamma$.}
\end{figure}

\begin{figure}[H]
\centering
\includegraphics[width=12cm, height=3cm]
{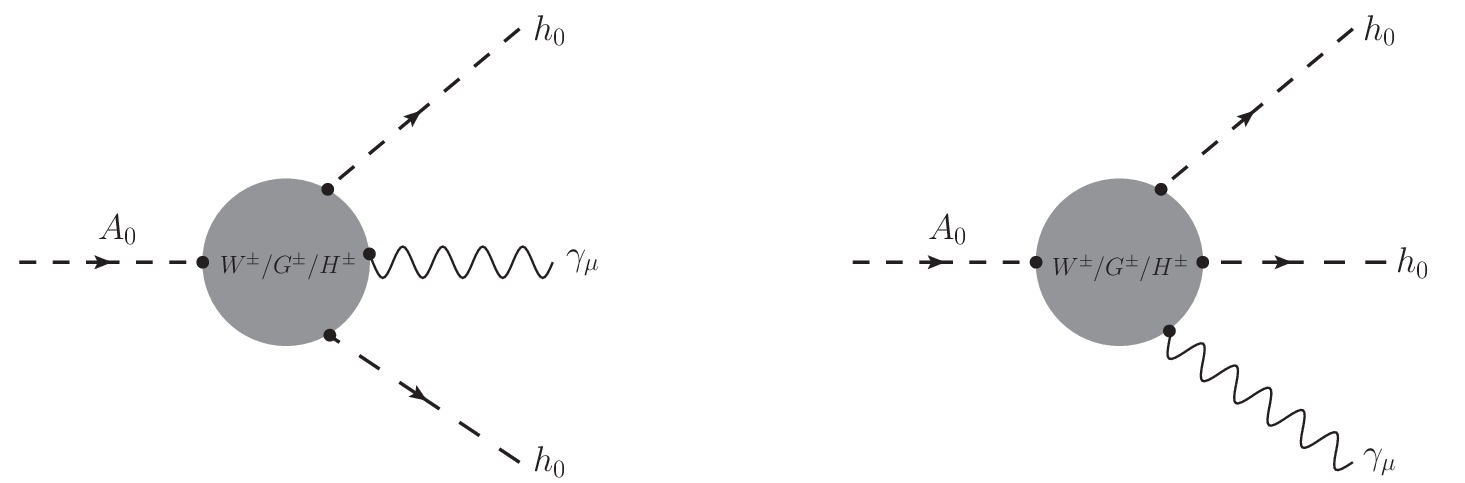}
\includegraphics[width=12cm, height=3cm]
{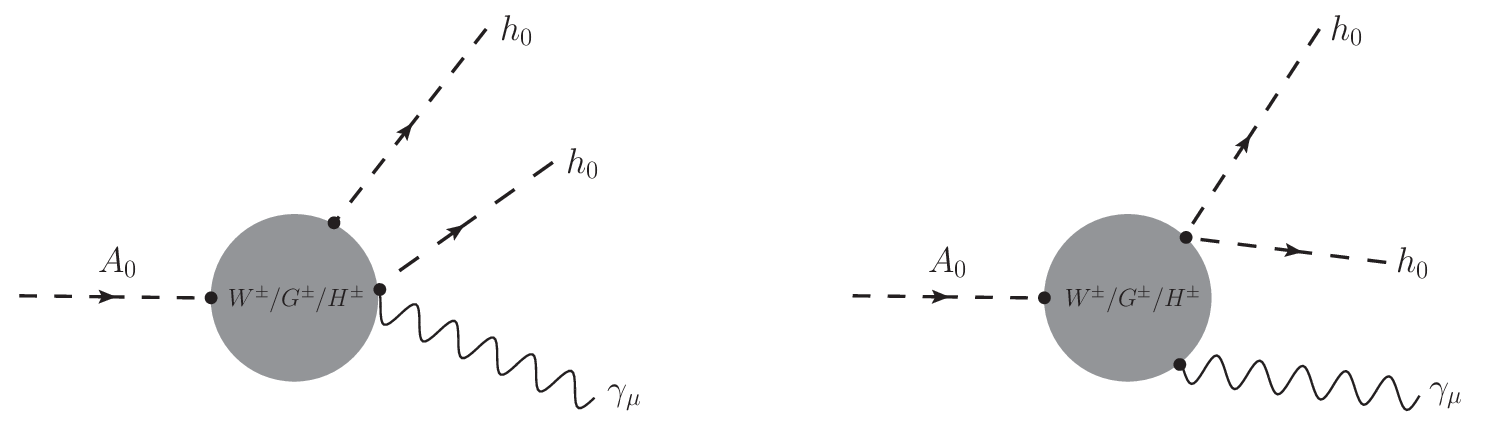}
\includegraphics[width=12cm, height=3cm]
{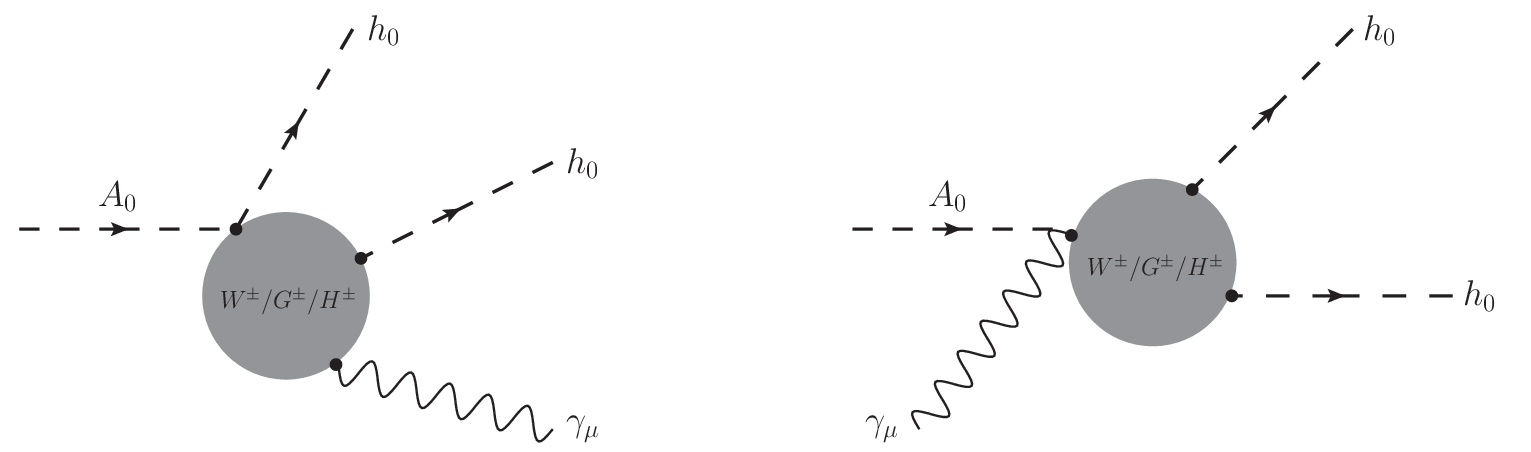}
\caption{
\label{BoxDiag}
One-loop four-point Feynman diagrams
contributing to the considered processes.
We only have charged Higgses and $W$ bosons 
together with Goldstone bosons in HF gauge
are considered to exchange in the loop.
}
\end{figure}

The detailed calculations for
one-loop induced for the rare 
decay process 
$A_0 \rightarrow h_0h_0\gamma$
in the THDM are discussed in this 
section. The computations are
handled with the help of the computer 
packages~{\tt FeynArts/FormCalc}
\cite{Hahn:2000kx}.
One-loop analytic formulas are
presented via scalar
PV-functions following the
standard output of the
programs~{\tt LoopTools/Collider}~\cite{Hahn:1998yk,Denner:2016kdg}.
First, one-loop Feynman
diagrams for the computed processes
are generated automatically by
using the package~{\tt FeynArts}
within the 't Hooft-Feynman gauge 
(HF). In this computation, 
we employ on-shell renormalization 
scheme developed in Refs.~\cite{Bohm:1986rj,
Hollik:1988ii,Aoki:1982ed} for
both fermion sector and gauge sector.
While the improved on-shell 
renormalization scheme is applied
for the scalar Higgs sectors 
following in Ref.~\cite{Kanemura:2017wtm}.
We list all one-loop
induced Feynman diagrams for
the channel 
$A_0\rightarrow h_0h_0\gamma$. 
They are categorized into several 
groups showing in the next
paragraphs. We first mention one-loop
Feynman diagrams with $A_0^*$-pole as 
plotted in Fig.~\ref{poleA0}.
These diagrams relating to off-shell 
decay process of $A_0^*
\rightarrow h_0\gamma$ 
in connection with the vertex
$A_0^*A_0h_0$.
The second classification
is included
all Feynman diagrams with 
$\phi^*$-poles
for $\phi^* = 
h_0^*,~H^*$ which are 
connected loop-induced processes
$\phi^* \rightarrow h_0\gamma$
with the vertex of
$\phi^*h_0h_0$, as plotted in 
Fig.~\ref{polePHI}. 
We next take into 
account all one-loop Feynman
diagrams with $Z^*$-pole as shown in
Fig.~\ref{Zpole}. In all the above-cases,
we also have the mixing of $Z$-$\gamma$ 
contributing to the considered 
process including for $A^*$-, 
$\phi^*$- and $Z^*$-poles 
diagrams. Finally, we have
one-loop four-point diagrams
taking into consideration 
in this computed channel.
For this group, we plot all one-loop
four-legs topologies as in
Fig.~\ref{BoxDiag}. We note that
all particles like fermion
$f$, vector bosons and scalar
particles are considered to 
exchange in the loop diagrams. 
Within 
the HF gauge, the Goldstone 
and Ghost particles are also 
propagated in the loop. 
As indicated in latter,
the fermion exchanging in the loop 
at several the above-groups
gives zero contribution. 
For this reason,
we skip showing fermion $f$ 
in the loop accordingly in 
these corresponding 
groups.

In general, one-loop
amplitude for the decay channel
$A_0(p) \rightarrow
h_0(k_1) h_0(k_2)\gamma_{\mu}(k_3)$
is given by following Lorentz 
structure as follows:
\begin{eqnarray}
\label{oneAmp}
i\mathcal{M}_{A_0
\rightarrow h_0h_0\gamma}
&=& \Big[
F_1 k_1^{\mu} + F_2 k_2^{\mu}\Big]
\varepsilon_{\mu}(k_3)
\\
\label{wardfactor}
&=&
F\Big[\frac{k_1^{\mu}}{k_1\cdot k_3}
-
\frac{k_2^{\mu}}{k_2\cdot k_3}
\Big]
\varepsilon_{\mu}(k_3),
\end{eqnarray}
where $\varepsilon_{\mu}(k_3)$ is the polarization vector of the external photon, three scalar factors $F$ and $F_{1,2}$ are loop contributions. The amplitude in Eq.~\eqref{oneAmp} follows the Ward identity relating to this on-shell  final photon state, i.e.,  this amplitude  will  equal to zero after  replacing  $\varepsilon_{\mu}(k_3) \rightarrow k_{3, \mu}$. Subsequently, we derive the relation for
$F_1,~F_2$ as follows:
\begin{eqnarray}
\label{wardIDD}
 F = (k_1\cdot k_3)F_1
 = -(k_2\cdot k_3) F_2,
\end{eqnarray}
resulting in Eq. ~\eqref{wardfactor}. More important,  the decay amplitude under consideration can be calculated via one of the two form factors $F_1$ or $F_2$. It means that we can collect form factor $F$ as one of the coefficients of $k_1^{\mu}$ (or $k_2^{\mu}$). One-loop form factor $F$ is expressed as functions of following kinematic
invariant masses:
\begin{eqnarray}
p^2 
&=& 
M_{A_0}^2,
\quad
k_1^2 =
k_2^2 = M_{h_0}^2,
\quad
k_3^2 =0,
\quad
k_{ij} = (k_i+k_j)^2,
\quad
\textrm{for}
\;
i,j =1,2,3.
\end{eqnarray}
We have further relation as
$k_{12}
+
k_{13}
+
k_{23}
= 2M_{h_0}^2
+ M_{A_0}^2$.
We are going to present
one-loop form factors
in the next paragraphs.
In general, one-loop 
form factors are decomposed 
into the form of
\begin{eqnarray}
\label{masterFORM}
F(M_{A_0}^2,M_{h_0}^2;
k_{12}, k_{13},\cdots )
&=&
\dfrac{g_{h_0A_0A_0}}
{
k_{13}
-
M_{A_0}^2
+
i
\Gamma_{A_0}
M_{A_0}
}
\cdot
F^{(\textrm{Trig})}_{A_0}
\\
&&
+
\sum
\limits_{
\phi=\{h_0^*, H^*\}
}
\dfrac{g_{\phi h_0h_0}}
{k_{12} - M_{\phi}^2
+ i\Gamma_{\phi}M_{\phi}}
\cdot
F^{(\textrm{Trig})}_{\phi}
\nonumber\\
&&
+
\sum
\limits_{ij=\{13, 23\}}
\dfrac{g_{A_0Zh_0}}
{
k_{ij}^2-M_{Z}^2
+ i\Gamma_{Z}M_{Z}
}
\cdot
F^{(\textrm{Trig})}_{Z, ij}
\nonumber\\
&&
+
\sum
\limits_{
N_S=\{S,SS,\cdots\}
}
F^{(\textrm{Box})}_{
(N_S, W)}
.
\nonumber
\end{eqnarray}
Where general couplings are given
\begin{eqnarray}
 g_{h_0A_0A_0} &=&
 \dfrac{-ie }{2M_Ws_{\beta}}
 \Big[
 s_{\beta-\alpha}(2M_{A_0}^2-M_{h_0}^2)
 + 
 \dfrac{c_{\beta+\alpha}}
 {s_{2\beta}} 
 \Big(
 2 M_{h_0}^2 
 - \dfrac{8M_W^2 s_W^2}{e^2 v^2}M^2 
 \Big)
 \Big],
 \\
 g_{h_0h_0h_0}&=&
 \dfrac{-3ie }{2M_W 
 s_W s_{2\beta}}
 \Big[
 (2c_{\alpha+\beta}+s_{2\alpha}
 s_{\beta-\alpha})M_{h_0}^2
 - 
 \dfrac{8M_W^2 s_W^2
 c_{\beta+\alpha}
  c^2_{\beta-\alpha}
 }{e^2 v^2}
 M^2 
 \Big)
 \Big],
 \\
g_{Hh_0h_0}&=&
-
\dfrac{
[s_{2\alpha}(3M^2-M_H^2-2M_{h_0}^2)
-M^2s_{2\beta}]
}{v\; s_{2\beta}}
\;
c_{\alpha-\beta}
,
\\
g_{A_0Zh_0}&=&
\left(
\dfrac{e}
{2c_Ws_W}
\right)
c_{\beta-\alpha}.
\end{eqnarray}

In the above equation, the first three  
form factors $F^{(\textrm{Trig})}_{A_0}$
and $F^{(\textrm{Trig})}_{\phi}$ 
($F^{(\textrm{Trig})}_{Z, ij}$)
correspond to the 
contributions from the 
$A_0^*$-pole and $\phi^*$-pole
($Z^*$-pole) diagrams, respectively. While 
the remaining one-loop form 
factors calculated from  
one-loop four-point diagrams
are decomposed into 
$F^{(\textrm{Box})}_{(S, W)}$,
$F^{(\textrm{Box})}_{(SS, W)}$, 
$F^{(\textrm{Box})}_{(SSS, W)}$
terms by
the numbers $N_S$ of charged  
Higgs ($S\equiv H^\pm$) 
in the loop. In the box diagrams
presenting in the computed processes,
the maximum number of charged Higgs
is $N_S=3$ in internal lines.
We first show the factors collected
from $A_0^*$-pole diagrams as shown in
Feynman diagrams in
Fig.~\ref{poleA0}. The resulting is
presented in terms of scalar
PV-functions as follows:
\begin{eqnarray}
\dfrac{
F^{(\textrm{Trig})}_{A_0}
}{
k_1\cdot k_3
}
&=&
-
\dfrac{i \, e^3}{(4\pi)^2}
\left(
\dfrac{
c_{\beta - \alpha}
}
{
2 M_W^2
s_W^2
}
\right)
\Bigg\{
M_W^2
\Big[
\big(
B_1 - B_0
\big)
[M_{h_0}^2; H^\pm, W]
-
\big(
5 B_0
+
B_1
\big)
[k_{13}; H^\pm, W]
\Big]
\nonumber\\
&&
+ M_W^2
\Big[
4 C_{00}
+
(M_{h_0}^2 + 3 k_{13}) C_{22}
\nonumber
\\
&&
\hspace{0.0cm}
+
(3 M_{h_0}^2 + k_{13}) C_{11}
+
4 (M_{h_0}^2 + k_{13}) C_{12}
\Big]
[M_{h_0}^2, 0, k_{13}; H^\pm, W, W]
\nonumber \\
&&
+
\Big[
M_W^2 (3 M_{h_0}^2+2 M_{H^\pm}^2
-M_{A_0}^2-4 M_W^2)
+
2 (M_{A_0}^2-M_{H^\pm}^2)
(M_{h_0}^2-M_{H^\pm}^2)
\Big]
\times
\nonumber\\
&& 
\times 
C_{0}[M_{h_0}^2, 0,
k_{13}; H^\pm, W, W]
\nonumber \\
&&
+
\Big[
M_{A_0}^2
(2 M_{h_0}^2-2 M_{H^\pm}^2+M_W^2)
-
2 M_{h_0}^2 (M_{H^\pm}^2-5 M_W^2)
\nonumber \\
&&\hspace{0.0cm}
+
2 M_{H^\pm}^2 (M_{H^\pm}^2-M_W^2)
-
M_W^2 k_{13}
\Big]
C_{1}[M_{h_0}^2, 0,
k_{13}; H^\pm, W, W]
\nonumber \\
&&
+
\Big[
M_W^2
(
M_{A_0}^2
-
2 M_{H^\pm}^2
+
4 M_{h_0}^2
+
5 k_{13}
)
+
2 (M_{A_0}^2-M_{H^\pm}^2)
(M_{h_0}^2-M_{H^\pm}^2)
\Big]
\times 
\nonumber \\
&&
\times 
C_{2}[M_{h_0}^2, 0,
k_{13}; H^\pm, W, W]
\nonumber \\
&&
+
\Big[
2 M_{A_0}^2 (M_{h_0}^2-M_{H^\pm}^2)
-
2 M_W^2 k_{13}
-
2 M_{H^\pm}^2
(M_{h_0}^2-M_{H^\pm}^2+M_W^2)
\Big]
\times 
\nonumber \\
&&
\times 
C_{2}[0, k_{13},
M_{h_0}^2; H^\pm, H^\pm, W]
\nonumber \\
&&
+
2 M_W^2
\Big[
2 C_{00}
+
(M_{h_0}^2-k_{13})
C_{12}
+
(M_{h_0}^2+k_{13})
C_{22}
\Big]
[0, k_{13},
M_{h_0}^2; H^\pm, H^\pm, W]
\Bigg\}
\nonumber\\
&&
-
\dfrac{ie^3}{(4\pi)^2}
\left(
\dfrac{
c_{\beta-\alpha}
}{2M_Z^2c_Ws_W}
\right)
\cdot
\Pi_{Z\gamma}[k_3^2=0, 
W^{\pm}, H^{\pm}],
\end{eqnarray}
where the two-point contribution relating to the $Z$-$\gamma$ mixing  is given in the 
following paragraph (as in Eq.~\eqref{mixingZA}). Because of the appearance of the on-shell photon state, this two-point contribution at $k_3^2=0$ is only contributed from the charged Higgs and $W$ bosons in the loop. In detail, the mixing is given by
\begin{eqnarray}
\label{mixingZA}
\Pi_{Z\gamma}[k_3^2=0, 
W^{\pm}, H^{\pm}]
&=&
\dfrac{2
(M_Z^2 - 6 M_W^2)
}{M_W^2}
\cot_W
\Big(
A_0[W]
-
2
B_{00}[0; W, W]
\Big)
\nonumber\\
&&
+
4 M_Z^2 \cot_W
\,
B_0[0; W, W]
\nonumber\
\\
&&
+4
\cot_{2W}
\Bigg(
-A_0[H^\pm]
+
2
B_{00}[0; H^\pm, H^\pm]
\Bigg).
\end{eqnarray}
In the analytic expressions, 
the scalar
one-loop coefficients 
$A_{ijk\cdots},~B_{ijk\cdots}
,~C_{ijk\cdots}$
(and $D_{ijk\cdots}$ appear in 
the later formulas)
are so-called as the scalar PV-functions
which are defined as in Ref.~\cite{Denner:1991kt}.
They are presented as standard output 
of both the packages {\tt LoopTools}
and {\tt Collider}. In this work, the  
PV-functions are presented
in the modified notations as:
\begin{eqnarray}
\{A,~B,~C,~D\}_{ijk\cdots}
[k_{ij}, \cdots;
M_{A}^2, M_{B}^2,M_{C}^2,
\cdots]
=\{A,~B,~C,~D\}_{ijk\cdots}
[k_{ij}, \cdots;
A,B,C, \cdots].
\end{eqnarray}
Where names of internal particles
$A,B,C, \cdots$
stand for their invariant masses 
$M_{A}^2, M_{B}^2,M_{C}^2,
\cdots$, respectively. 

One-loop form factors contributing
from the diagrams with $\phi^*$-poles
for $\phi=h_0,~H$ 
are next collected. One-loop Feynman
diagrams for this group are plotted
in Fig.~\ref{polePHI}.
The corresponding factors are
given in the form as follows:
\begin{eqnarray}
\dfrac{
F^{(\textrm{Trig})}_{\phi}
}
{k_1\cdot k_3}
&=&
\dfrac{i \, e^2}{(4\pi)^2}
\cdot
\left(
\dfrac{
g_{\phi \, H^\pm W^\mp}
}
{M_W^2 s_W}
\right)
\Bigg\{
M_W^2
\big(
B_1 - B_0
\big)
[M_{A_0}^2; H^\pm, W]
-
M_W^2
\big(
B_1 + 5 B_0
\big)
[k_{12}; H^\pm, W]
\nonumber \\
&&
+
M_W^2
\Big[
4 C_{00}
+
(3 M_{A_0}^2+k_{12}) C_{11}
\nonumber
\\
&&
\hspace{0.0cm}
+
4 (M_{A_0}^2+k_{12}) C_{12}
+
(M_{A_0}^2+3 k_{12}) C_{22}
\Big]
[M_{A_0}^2, 0, k_{12}; H^\pm, W, W]
\nonumber \\
&&
+
\Big[
M_W^2 (3 M_{A_0}^2-M_{\phi}^2
+2 M_{H^\pm}^2-4 M_W^2)
+
2 (M_{A_0}^2-M_{H^\pm}^2)
(M_{\phi}^2-M_{H^\pm}^2)
\Big]
\times
\nonumber\\
&&
\times
C_0[M_{A_0}^2, 0, k_{12}; H^\pm, W, W]
\nonumber \\
&&
+
\Big[
2 M_{A_0}^2 (M_{\phi}^2-M_{H^\pm}^2+5 M_W^2)
- M_{\phi}^2 (2 M_{H^\pm}^2 - M_W^2)
\nonumber \\
&&\hspace{4.0cm}
- 2 M_{H^\pm}^2 (M_W^2 - M_{H^\pm}^2)
- M_W^2 k_{12}
\Big]
C_1[M_{A_0}^2, 0, k_{12}; H^\pm, W, W]
\nonumber \\
&&
+
\Big[
2 (M_{A_0}^2-M_{H^\pm}^2)
(M_{\phi}^2-M_{H^\pm}^2)
+
M_W^2 (
4 M_{A_0}^2
-
2 M_{H^\pm}^2
+
M_{\phi}^2
+
5 k_{12}
)
\Big]
\times
\nonumber\\
&&
\times
C_2[M_{A_0}^2, 0, k_{12}; H^\pm, W, W]
\nonumber \\
&&
+
2 M_W^2
\Big[
(M_{A_0}^2 - k_{12}) C_{12}
+
(k_{12} + M_{A_0}^2) C_{22}
+
2 C_{00}
\Big]
[0, k_{12}, M_{A_0}^2;
H^\pm, H^\pm, W]
\nonumber \\
&&
+
\Big[
2 (M_{A_0}^2 - M_{H^\pm}^2)
(M_{\phi}^2 - M_{H^\pm}^2)
-
2 M_W^2 (M_{H^\pm}^2+k_{12})
\Big]
C_2[0, k_{12}, M_{A_0}^2;
H^\pm, H^\pm, W]
\Bigg\}
\nonumber\\
&&
+
\dfrac{ie^2}{(4\pi)^2}
\left(
\dfrac{
g_{A_0Z\phi}
}{M_Z^2s_{2W}}
\right)
\cdot
\Pi_{Z\gamma}[k_3^2=0,W^\pm, H^\pm].
\end{eqnarray}
Where the general couplings are
shown as
$g_{A_0Z\phi} =
\frac{e\;c_{\beta-\alpha}}
{2c_Ws_W}~\left(
-\frac{e\;s_{\beta-\alpha}}
{2c_Ws_W}\right)$
for $\phi =h_0~(H)$, 
respectively.
Other general couplings 
are presented 
as $g_{\phi \, H^\pm W^\mp}
= \frac{-ie\;c_{\beta-\alpha}}
{2s_W}~\left(\frac{ie\;s_{\beta-\alpha}}
{2s_W} \right)$
for $\phi =h_0~(H)$, correspondingly.
In further, the mixing $Z$-$\gamma$ 
is also given by Eq.~\eqref{mixingZA}.

We next consider one-loop triangle 
diagrams with $Z$-pole including 
the mixing of $Z$-$\gamma$ attributing
to one-loop form factors. All internal
particles such as fermion, vector
bosons, scalar particles are considered
exchanging in the loop. These diagrams
are plotted in Fig.~\ref{Zpole}.
There are two form factors relating
to the contributions of $Z$-pole 
giving in the above equation like
$F^{(\textrm{Trig})}_{Z, 13}$ and 
$F^{(\textrm{Trig})}_{Z, 23}$. 
Different from the previous 
form factors, both factors for 
$Z$-pole have non-zero contributions 
from fermion 
$f$ exchanging in the loop.
We show the analytic results for 
one-loop form factors 
$F^{(\textrm{Trig})}_{Z, ij}$ 
as follows:
\begin{eqnarray}
F^{(\textrm{Trig})}_{Z, ij}
&=&
\sum\limits_{P=\{f,W,H^\pm \}}
F^{(\textrm{Trig},P)}_{Z, ij}
+
F^{(Z\textrm{-}\gamma)}_{Z, ij}
\end{eqnarray}
for $ij =\{13,~23\}$. 
Where each factor is 
given accordingly:
\begin{eqnarray}
\label{zfermion13}
\dfrac{
F^{(\textrm{Trig},f)}_{Z, 13}
}{k_1\cdot k_3}
&=&-
\sum\limits_{f}
\dfrac{e\;Q_f\; N_C^f
m_f^2}{2\pi^2}
g_{h_0 f \bar{f} } \times
g_{Z f \bar{f} }
\times
\\
&&
\times
\Bigg\{
B_0[k_{13}; f, f]
+
\Big[
2
(3 M_{h_0}^2
-
2 k_{12}
-
k_{13})
C_{11}
-
2
(3 M_{h_0}^2
-
2 k_{23}
-
k_{13})
C_{12}
\nonumber \\
&&
+
(M_{h_0}^2
-
2 k_{12}
+
2 k_{23}
-
k_{13})
C_1
-
(M_{h_0}^2-k_{23})
C_0
-
4 C_{00}
\Big]
[M_{h_0}^2, k_{13}, 0;
f, f, f]
\Bigg\}.
\nonumber
\end{eqnarray}
Here $N_C^f$ stands for number for 
color of the corresponding fermion $f$. 
It will be 1 for leptons and 3 for quarks.
The general coupling $g_{h_0 f \bar{f} }$ 
is given in Table~\ref{Z2-assignment} 
in the section $2$. While 
$g_{Z f \bar{f} }$ are couplings of 
$Z$ with fermion-pair which are 
taken in the SM.
Other factors are presented as:
\begin{eqnarray}
\dfrac{
F^{(\textrm{Trig}, W)}_{Z, 13}
}{k_1\cdot k_3}
&=&
\dfrac{e^3}{(4\pi)^2}
\dfrac{
c_W\;s_{\beta-\alpha}
}{2 M_W^2\;s_W^2}
\Bigg\{
2 M_W^2 (M_W^2-M_Z^2)
B_0[0; W, W]
\\
&&
+
\Big\{
M_W^2
(M_{h_0}^2 - M_{A_0}^2)
B_1
\nonumber\\
&&
+
\Big[
M_W^2 (M_{A_0}^2+12 M_W^2)
+
M_{h_0}^2 (M_W^2 - M_Z^2)
\Big]
B_0
\Big\}
[M_{h_0}^2; W, W]
\nonumber \\
&&
+
M_W^2
\Big[
3 (M_{h_0}^2 - M_{A_0}^2)
(B_1 - B_0)
-
2 (M_W^2 + 2 M_Z^2)
B_0
\Big]
[k_{13}; W, W]
\nonumber \\
&&
-
4 \Big[
2 M_W^2
(M_{A_0}^2 + 6 M_W^2)
-
M_Z^2
(M_{h_0}^2 + 2 M_W^2)
\Big]
C_{00}[M_{h_0}^2, 
k_{13}, 0; W, W, W]
\nonumber \\
&&
+
2 M_W^2
\Big[
3 M_W^2
(M_{A_0}^2-M_{h_0}^2-M_Z^2)
+
2 M_{h_0}^2
(M_{A_0}^2
-
M_{h_0}^2
-
4 M_W^2)
\nonumber\\
&&
+
k_{12}
(M_Z^2 - 3 M_W^2)
\nonumber \\
&&\hspace{0cm}
+
k_{13}
(M_{h_0}^2 - M_{A_0}^2+M_Z^2)
+
k_{23}
(13 M_W^2-3 M_Z^2)
\Big]
C_{0}[M_{h_0}^2, 
k_{13}, 0; W, W, W]
\nonumber \\
&&
+
\Big\{
M_W^2
\Big[
M_{h_0}^2
(
11 M_{h_0}^2
-
5 M_{A_0}^2
+
60 M_W^2
)
\nonumber \\
&&
\hspace{0cm}
+
k_{13}
(
M_{h_0}^2 - 3 M_{A_0}^2 - 20 M_W^2
)
-
4 k_{12} (M_{h_0}^2+10 M_W^2)
\Big]
\nonumber \\
&&\hspace{0cm}
+
2 (M_{h_0}^2+2 M_W^2)
(M_W^2-M_Z^2)
(3 M_{h_0}^2-2 k_{12}-k_{13})
\Big\}
C_{11}[M_{h_0}^2, k_{13}, 
0; W, W, W]
\nonumber \\
&&
- 2
\Big[
M_{h_0}^2 M_W^2
(
2 M_{A_0}^2
+
M_{h_0}^2
+
30 M_W^2
)
\nonumber \\
&&\hspace{0cm}
-
M_W^2 k_{13}
(
2 M_{A_0}^2
-
M_{h_0}^2
+
10 M_W^2
)
-
2 M_W^2 k_{23}
(M_{h_0}^2 + 10 M_W^2)
\nonumber \\
&&\hspace{0cm}
+
(M_{h_0}^2+2 M_W^2)
(M_W^2-M_Z^2)
(3 M_{h_0}^2-k_{13}-2 k_{23})
\Big]
C_{12}[M_{h_0}^2,
k_{13}, 0; W, W, W]
\nonumber \\
&&
+
\Big\{
M_{h_0}^2 M_W^2
( 4 M_{A_0}^2
-
4 M_{h_0}^2
-
4 M_W^2
-
9 M_Z^2 )
+
M_W^2 k_{13}
(
M_{h_0}^2
+
M_Z^2
-
M_{A_0}^2
)
\nonumber \\
&&
+
k_{12}
\Big[
(2 M_{h_0}^2+5 M_W^2)
(M_Z^2-M_W^2)
-
M_W^2
( 19 M_W^2
+
2 M_{h_0}^2 )
\Big]
\nonumber \\
&&
+
k_{23}
\Big[
(2 M_{h_0}^2+3 M_W^2)
(M_W^2-M_Z^2)
+
M_W^2
(
2 M_{h_0}^2
+
21 M_W^2
)
\Big]
\Big\}
\times
\nonumber\\
&&
\times
C_{1}[M_{h_0}^2, k_{13}, 0; W, W, W]
\nonumber \\
&&
+
\Big[
2 M_W^2
(
M_{h_0}^2
-
k_{13}
)
(
2 M_{A_0}^2
-
2 M_{h_0}^2
-
M_W^2
-
M_Z^2
)
\Big]
C_{2}[M_{h_0}^2, k_{13}, 0; W, W, W]
\Bigg\},
\nonumber \\
& \nonumber\\
\dfrac{
F^{(\textrm{Trig}, H^{\pm})}_{Z, 13}
}{k_1\cdot k_3}
&=&
-
\dfrac{ie^2}{(4\pi)^2}
\dfrac{\cot_{2W}
}{2 M_W^4 }
\cdot
g_{h_0 H^\pm H^\mp}
\times
\\
&&
\times
\Bigg\{
- 4 M_W^4
B_0[M_{h_0}^2; H^\pm, H^\pm]
+
8 M_W^4
\Big[
2 C_{00}
-
(3 M_{h_0}^2-2 k_{12}- k_{13})
C_{11}
\nonumber \\
&&
+
(3 M_{h_0}^2- k_{13}-2 k_{23})
C_{12}
+
(k_{12}- k_{23})
C_{1}
\Big]
[M_{h_0}^2, k_{13}, 0;
H^\pm, H^\pm, H^\pm]
\Bigg\},
\nonumber\\
& \nonumber\\
\dfrac{
F^{(Z\textrm{-}\gamma)}_{Z, 13}
}{k_1\cdot k_3}
&=&-
\dfrac{e^3\;M_W\;
s_{\beta-\alpha}}
{(4\pi)^2\; s_W
\; c_W^2}
\left(
\dfrac{M_{h_0}^2
-
M_{A_0}^2
+
M_Z^2}
{
4M_Z^4}
\right)
\cdot
\Pi_{Z\gamma}
[k_3^2=0, W^\pm, H^\pm].
\end{eqnarray}
Other factors collected 
from $Z$-pole diagrams
are given by
\begin{eqnarray}
\label{zfermion23}
\dfrac{
F^{(\textrm{Trig},f)}_{Z,23}
}
{
k_1\cdot k_3
}
&=&-
\sum\limits_{f}
\dfrac{e Q_f N_C^f \times
m_f^2}{2\pi^2}
g_{h_0 f \bar{f} } \times
g_{Z f \bar{f} }\times
\\
&&
\times
\Bigg\{
2 B_0[k_{23}; f, f]
+
\Big[
(M_{h_0}^2-k_{23})
\big(
C_0
+
2 C_1
\big)
-
8 C_{00}
\Big]
[M_{h_0}^2, k_{23}, 0;
f, f, f]
\Bigg\},
\nonumber\\
\dfrac{
F^{(\textrm{Trig},W)}_{Z,23}
}
{
k_1\cdot k_3
}
&=&
\dfrac{e^3}{(4\pi)^2}
\dfrac{c_W s_{\beta-\alpha}}
{M_W^3\; s_W^2}
\Bigg\{
2 M_W^2
(M_W^2-M_Z^2)
B_0[0; W, W]
\\
&&
+
\Big[
M_{h_0}^2
(2 M_W^2-M_Z^2)
+
12 M_W^4
\Big]
B_0[M_{h_0}^2; W, W]
\nonumber\\
&&
-
2 M_W^2
\big(
M_W^2
+
2 M_Z^2
\big)
B_0[k_{23}; W, W]
\nonumber \\
&&
-4
\Big[
(M_{h_0}^2+2 M_W^2)
(2 M_W^2-M_Z^2)
+
8 M_W^4
\Big]
C_{00}[M_{h_0}^2,
k_{23}, 0; W, W, W]
\nonumber \\
&&
+
2 M_W^2
\Big[
M_{h_0}^2
(7 M_W^2-3 M_Z^2)
+
k_{23}
(2 M_Z^2 - 5 M_W^2)
-
3 M_Z^2 M_W^2
\Big]
\times 
\nonumber\\
&&
\times 
C_{0}[M_{h_0}^2,
k_{23}, 0; W, W, W]
\nonumber \\
&&
-
2 M_W^2
\Big[
M_{h_0}^2
(
2 M_W^2
+
3 M_Z^2
)
-
M_Z^2 k_{23}
\Big]
C_{1}[M_{h_0}^2,
k_{23}, 0; W, W, W]
\nonumber \\
&&
-
\Big[
2 M_W^2
(M_{h_0}^2 - k_{23})
(M_W^2 + M_Z^2)
\Big]
C_{2}[M_{h_0}^2,
k_{23}, 0; W, W, W]
\Bigg\},
\nonumber\\
\dfrac{
F^{(\textrm{Trig},H^{\pm})}_{Z,23}
}
{
k_1\cdot k_3
}
&=&
-
\dfrac{ie^2}{(4\pi)^2}
\left(
\dfrac{
\cot_{2W}
}{M_W^4}
\right)
\cdot
g_{h_0 H^\pm H^\mp}
\times
\\
&&
\times
\Bigg\{
-
4 M_W^4
B_0[M_{h_0}^2; H^\pm, H^\pm]
+
16 M_W^4
C_{00}[M_{h_0}^2, k_{23}, 0;
H^\pm, H^\pm, H^\pm]
\Bigg\},
\nonumber\\
\dfrac{
F^{(Z\textrm{-}\gamma)}_{Z,23}
}
{
k_1\cdot k_3
}
&=&
-
\dfrac{e^3\;M_W\;
s_{\beta-\alpha}}
{(4\pi)^2\; s_W
\; c_W^2}
\left(
\dfrac{1}{ M_Z^2}
\right)
\cdot
\Pi_{Z\gamma}[k_3^2=0,W^\pm, H^\pm].
\end{eqnarray}
Where the general coupling given
in the above equations
\begin{eqnarray}
\label{hHH}
g_{h_0 H^\pm H^\mp}=
-
\dfrac{c_{\alpha+\beta}
(4M^2-3M_h^2 - 2M_{H^\pm}^2)
+(2M_{H^\pm}^2-M_h^2)
c_{(\alpha-3\beta)}}
{2vs_{2\beta}}
\end{eqnarray}
is taken as in Ref.~\cite{Phan:2024jbx}.
Here the mixing of $Z$-$\gamma$ is also
used the equation~\eqref{mixingZA}.

We turn our attention to the contributions 
from one-loop four-point Feynman diagrams.
All the diagrams plotted within the HF gauge 
are shown in the below paragraphs.
We confirm that the contributions 
from fermion $f$ in the loop of all 
one-loop box diagrams are vanished
in this case.
One-loop form factors
$F^{(\textrm{Box})
}_{(S\cdots, W)}$
are collected from one-loop
four-point diagrams which
are divided into sub-factors 
following number
of charged Higges in the loop.
We first present the factors
deriving from the diagrams
with one charged Higgs in
the loop. The factors are
taken in the form of
\begin{eqnarray}
\dfrac{
F^{(\textrm{Box})}_{
(S,W)
}
}
{k_1\cdot k_3}
&=&
\frac{e^2}{8\pi^2}
\left(
\frac{
g_{h_0 h_0 H^\pm G^\mp}
}{2M_Ws_W}
\right)
\cdot
G_0
+
\frac{e^4}{32\pi^2}
\left(
\frac{s_{2(\beta-\alpha)}}
{M_W^3 s_W s^2_{2W}}
\right)
\cdot
\sum
\limits_{i = 1}^{5}
G_i.
\end{eqnarray}
The general coupling
$g_{h_0 h_0 H^\pm G^\mp}$
is taken into account in
the above equation
(this coupling is 
collected as 
in Ref.~\cite{Phan:2024jbx})
as follows
\begin{eqnarray}
\label{hhHG}
g_{h_0 h_0 H^\pm G^\mp}
&=&
\frac{ e^2 \, c_{\beta - \alpha} }
{
4 M_W^2 s_W^2 s_{2\beta}
}
\Big\{
2 c_{\alpha + \beta} 
(M_{h_0}^2 - M^2)
+
s_{\beta - \alpha} 
\Big[
s_{2\alpha} (M_{h_0}^2 - M_{H_0}^2)
-
2 s_{2\beta} (M_{H^\pm}^2 - M^2)
\Big]
\Big\}.
\nonumber\\
\end{eqnarray}
All factors $G_i$
for $i=0, 1, \cdots, 5$
are listed in terms of
PV-functions as follows
\begin{eqnarray}
\dfrac{G_0}{M_W^2}
&=&
\dfrac{
[2 M_{H^\pm}^2
- 2 M_{A_0}^2 + M_W^2]
}{M_W^2}
C_0[M_{A_0}^2, 0, k_{12};
H^\pm, W, W]
\\
&&
+
\dfrac{
[2 M_{H^\pm}^2
- 2 M_{A_0}^2 - M_W^2]
}
{M_W^2}
\big( C_1 + C_2 \big)
[M_{A_0}^2, 0, k_{12};
H^\pm, W, W].
\nonumber
\end{eqnarray}
Another factor is given by
\begin{eqnarray}
\dfrac{G_1}
{M_W^2}
&=&
\Big[
2 M_W^2 C_0
-
(M_{h_0}^2
- M_{H^\pm}^2
+ 2 M_W^2)
\big(C_1 + 2 C_2 \big)
\Big]
[M_{h_0}^2, M_{h_0}^2, k_{12};
H^\pm, W, W]
\nonumber \\
&&
-
\Big[
2
(M_{A_0}^2
-M_{H^\pm}^2
-M_W^2)
C_0
+
(M_{A_0}^2
-M_{H^\pm}^2
+2 M_W^2)
C_1
\Big]
[M_{A_0}^2,M_{h_0}^2,
k_{23}; H^\pm, W, W]
\nonumber 
\\
&&
-
\Big[
(M_{A_0}^2-M_{H^\pm}^2-2 M_W^2)
C_0
\nonumber \\
&&
+
(M_{A_0}^2-M_{H^\pm}^2+2 M_W^2)
\big(C_1 + C_2\big)
\Big]
[M_{A_0}^2,M_{h_0}^2,
k_{13}; H^\pm, W, W]
\nonumber \\
&&
+
\big(
M_{A_0}^2+M_{h_0}^2-2 M_{H^\pm}^2
\big)
\big(
C_0
-
C_1
-
C_2
\big)
[M_{h_0}^2, k_{23},
M_{A_0}^2; H^\pm, W, W]
\nonumber \\
&&
+
\Big[
(M_{h_0}^2-M_{H^\pm}^2)
C_0
-
(M_{A_0}^2
+M_{h_0}^2
-2 M_{H^\pm}^2)
C_2
\Big]
[M_{h_0}^2, k_{13}, M_{A_0}^2;
H^\pm, W, W]
\nonumber \\
&&
-
2 \big(M_{A_0}^2
-M_{H^\pm}^2-M_W^2 \big)
C_0[k_{23}, 0,
M_{h_0}^2; H^\pm, W, W]
\nonumber \\
&&
+
\Big[
(2 M_{A_0}^2+4 M_{h_0}^2
-6 M_{H^\pm}^2+5 M_W^2)
C_0
\nonumber \\
&&
+
(2 M_{A_0}^2
-2 M_{H^\pm}^2
+M_W^2)
\big(
C_1
+
C_2
\big)
\Big]
[M_{h_0}^2, 0,
k_{13}; W, W, W]
\nonumber \\
&&
+
2
\big(
M_{h_0}^2
+M_{A_0}^2
-2 M_{H^\pm}^2
+6 M_W^2
\big)
C_0[M_{h_0}^2, 0,
k_{23}; W, W, W]
\nonumber \\
&&\hspace{0.0cm}
+
\dfrac{
(M_{A_0}^2-M_{H^\pm}^2)
}{M_W^2}
\Big[
(-2 M_{h_0}^2+2 M_{H^\pm}^2+M_W^2)
C_0
\nonumber\\
&&
- (2 M_{h_0}^2-2 M_{H^\pm}^2+M_W^2)
\big(C_1 + C_2 \big)
\Big]
[M_{h_0}^2, 0, k_{13}; H^\pm, W, W],
\end{eqnarray}
and 
\begin{eqnarray}
\dfrac{G_2}
{M_W^2}
&=&
\Bigg\{
(-2 M_{A_0}^2+ 2 M_{H^\pm}^2-8 M_W^2)D_{00}
+
\Big[
-2 M_{h_0}^2 (-2 M_{h_0}^2 + M_W^2+k_{12})
\nonumber \\
&&
+
k_{12}
(M_{H^\pm}^2 - 2 M_W^2)
+
M_{A_0}^2
(k_{12}-4 M_{h_0}^2)
-
2 M_W^2 k_{13}
\Big]
D_{11}
\nonumber \\
&&
+
\Big[
(M_{A_0}^2-8 M_{h_0}^2+M_{H^\pm}^2
-2 M_W^2+k_{12}-2 k_{13}) M_{A_0}^2
-
6 M_W^2 k_{13}
\nonumber \\
&&
+
k_{12}
(M_{H^\pm}^2
-
2 M_W^2)
-
2 M_{h_0}^2
(-3 M_{h_0}^2+M_W^2+k_{12}-k_{13})
\Big]
D_{12}
\nonumber \\
&&
+
\Big[
(M_{A_0}^2+M_{H^\pm}^2-4 M_W^2
-k_{12}-2 k_{13}) M_{A_0}^2
+
k_{12}
(M_{H^\pm}^2
-
2 M_W^2)
\nonumber \\
&&
-
2 M_W^2 k_{13}
+
2 M_{h_0}^2
(-M_{h_0}^2+M_W^2+k_{13})
\Big]
D_{33}
\nonumber \\
&&
+
\Big[
-
(4 M_{h_0}^2-2 M_{A_0}^2
-2 M_{H^\pm}^2
+6 M_W^2+k_{12}+4 k_{13})
M_{A_0}^2
\nonumber \\
&&
+
k_{12}
(M_{H^\pm}^2 - 2 M_W^2)
-
6 M_W^2 k_{13}
+
2 M_{h_0}^2 (M_W^2+2 k_{13})
\Big]
D_{23}
\nonumber \\
&&
+
\Big[
(-4 M_{h_0}^2+M_{A_0}^2
+M_{H^\pm}^2-2 M_W^2
- 2 k_{13}) M_{A_0}^2
\nonumber\\
&&
+
2 M_{h_0}^2 (M_{h_0}^2 + k_{13})
-
4 M_W^2 k_{13}
\Big]
D_{22}
\nonumber \\
&&
+
\Big[
(-4 M_{h_0}^2+M_{A_0}^2
+M_{H^\pm}^2
-4 M_W^2 -2 k_{13}) M_{A_0}^2
+
2 M_{h_0}^2
(M_{h_0}^2+k_{13}-k_{12})
\nonumber \\
&&
+
2 M_{H^\pm}^2 k_{12}
-
4 M_W^2 (k_{12}+k_{13})
\Big]
D_{13}
\Bigg\}
[M_{h_0}^2, 0, M_{h_0}^2, M_{A_0}^2;
k_{13}, k_{23};
H^\pm, W, W, W]
\nonumber
\end{eqnarray}
\begin{eqnarray}
&&
+
\dfrac{1}
{M_W^2}
\Bigg\{
\Big[
M_{h_0}^2
(-2 M_{h_0}^2+2 M_{H^\pm}^2+M_W^2)
+
M_W^2 (4 M_{H^\pm}^2-
M_{A_0}^2 + M_W^2 +k_{13})
\Big]
M_{A_0}^2
\nonumber \\
&&
+
2 M_{h_0}^4 (M_{H^\pm}^2+M_W^2)
+
M_{H^\pm}^2 M_W^2
(-5 M_{H^\pm}^2+5 M_W^2+k_{13})
\nonumber \\
&&
-
M_{h_0}^2
\Big[
(2 M_{H^\pm}^2
+ M_W^2)
M_{H^\pm}^2
+2 M_W^2 k_{13}
\Big]
\Bigg\}
D_0[M_{h_0}^2, 0,
M_{h_0}^2, M_{A_0}^2;
k_{13}, k_{23};
H^\pm, W, W, W]
\nonumber \\
&&
+
\dfrac{1}
{M_W^2}
\Bigg\{
- M_{A_0}^2
\Big[
M_{h_0}^2
(2 M_{h_0}^2
-2 M_{H^\pm}^2
+M_W^2)
+
M_W^2
(- M_{A_0}^2
+M_W^2+k_{12}
+k_{13})
\Big]
\nonumber \\
&&
+
2 M_{h_0}^4
(M_{H^\pm}^2
+3 M_W^2)
+
M_W^2
\Big[
3 M_{H^\pm}^2
(- M_{H^\pm}^2
+M_W^2
+k_{12}
+k_{13})
+
2 M_W^2
(k_{12}+k_{13})
\Big]
\nonumber \\
&&
-
M_{h_0}^2
\Big[
(2 M_{H^\pm}^2
+3 M_W^2)
M_{H^\pm}^2
+2 M_W^2
(3 M_W^2+k_{12}
+k_{13})
\Big]
\Bigg\}
\times
\nonumber\\
&&
\times
D_1
[M_{h_0}^2, 0,
M_{h_0}^2, M_{A_0}^2;
k_{13}, k_{23};
H^\pm, W, W, W]
\nonumber \\
&&
+
\dfrac{1}{M_W^2}
\Bigg\{
-
M_{A_0}^2
\Big[
(2 M_{h_0}^2+M_W^2
-2 M_{H^\pm}^2)
M_{h_0}^2
+M_W^2 (M_{H^\pm}^2
+3 M_W^2+4 k_{12}
+3 k_{13})
\Big]
\nonumber \\
&&
+
2 M_{h_0}^4
(M_{H^\pm}^2+2 M_W^2)
-
M_{h_0}^2
\Big[
M_{H^\pm}^2
(2 M_{H^\pm}^2 + 3 M_W^2)
+
4 M_W^4
\Big]
\nonumber \\
&&
+
M_W^2
\Big[
4 M_{A_0}^4
+
M_{H^\pm}^2
(-3 M_{H^\pm}^2
+3 M_W^2
+4 k_{12}
+3 k_{13})
+
4 M_W^2 k_{12}
\Big]
\Bigg\}\times
\nonumber\\
&&
\times
D_2[M_{h_0}^2, 0,
M_{h_0}^2, M_{A_0}^2;
k_{13}, k_{23};
H^\pm, W, W, W]
\nonumber\\
&&
+
\dfrac{1}{M_W^2}
\Bigg\{
-
M_{A_0}^2
\Big[
M_W^2 (M_{H^\pm}^2+5 M_W^2
+3 k_{12}+ 3 k_{13})
-
4 M_W^2 M_{A_0}^2
\nonumber \\
&&
-
M_{h_0}^2
(2 M_{H^\pm}^2
-2 M_{h_0}^2+3 M_W^2)
\Big]
-
M_{h_0}^2
\Big[
M_{H^\pm}^2
(2 M_{H^\pm}^2
-
2 M_{h_0}^2
+
3 M_W^2)
+
2 M_W^4
\Big]
\nonumber\\
&&
+
M_W^2
\Big[
3 M_{H^\pm}^2
(M_W^2- M_{H^\pm}^2
+k_{12}+k_{13})
+2 M_W^2
(k_{12}+k_{13})
\Big]
\Bigg\}
\times 
\nonumber \\
&&
\times 
D_3
[M_{h_0}^2, 0,
M_{h_0}^2, 
M_{A_0}^2;
k_{13}, k_{23};
H^\pm, W, W, W].
\end{eqnarray}
Next, coefficient factors 
$G_3$, $G_4$ and $G_5$ are 
expressed as follows:
\begin{eqnarray}
\dfrac{G_3}
{M_W^2}
&=&
\Bigg\{
(
-2 M_{h_0}^2
+ 2 M_{H^\pm}^2
-8 M_W^2
)
D_{00}
\nonumber \\
&&
-
\Big[
2 k_{12} (M_{h_0}^2+2 M_W^2)
+
M_{A_0}^2
(
M_{h_0}^2
- M_{H^\pm}^2
+2 M_W^2
-2 k_{12}
)
\Big]
D_{22}
\nonumber \\
&&
+
\Big[
(3 M_{h_0}^2+M_{H^\pm}^2
-2 M_W^2-2 k_{12}-k_{13})
M_{h_0}^2
\nonumber \\
&&
-
2 M_W^2 k_{12}
+
M_{A_0}^2
(2 k_{12}-4 M_{h_0}^2)
+
(M_{H^\pm}^2
-
2 M_W^2) k_{13}
\Big]
D_{11}
\nonumber \\
&&
+
\Big[
(3 M_{h_0}^2+M_{H^\pm}^2
-2 M_W^2-4 k_{12}-k_{13}) M_{h_0}^2
+
(M_{H^\pm}^2
-
2 M_W^2) k_{13}
\nonumber \\
&&
-
6 M_W^2 k_{12}
+M_{A_0}^2 (-5 M_{h_0}^2
+M_{H^\pm}^2-2 M_W^2+4 k_{12})
\Big]
D_{12}
\nonumber \\
&&
-
\Big[
2 M_{A_0}^2 (- M_{A_0}^2
+M_{h_0}^2-M_{H^\pm}^2
+3 M_W^2-k_{12}+k_{13})
+
(2 M_W^2
-
M_{H^\pm}^2) k_{13}
\nonumber \\
&&
+6 M_W^2 k_{12}
+
M_{h_0}^2 (M_{H^\pm}^2+
M_{h_0}^2-2 M_W^2+2 k_{12}-k_{13})
\Big]
D_{23}
\nonumber \\
&&
+
\Big[
(-5 M_{h_0}^2
+2 M_{A_0}^2+M_{H^\pm}^2
-4 M_W^2+2 k_{12}-2 k_{13}) M_{A_0}^2
\nonumber \\
&&
+
2 M_{h_0}^2
(M_{h_0}^2 - k_{12})
+
2 k_{13}
(M_{H^\pm}^2
-2 M_W^2)
-
4 M_W^2 k_{12}
\Big]
D_{13}
\nonumber \\
&&
-
\Big[
(-2 M_{A_0}^2+M_{h_0}^2
-M_{H^\pm}^2+4 M_W^2
+2 k_{13}) M_{A_0}^2
+
2 M_W^2 k_{12}
+
(2 M_W^2 - M_{H^\pm}^2) k_{13}
\nonumber \\
&&
+
M_{h_0}^2
(
M_{H^\pm}^2
+M_{h_0}^2
-2 M_W^2-k_{13}
)
\Big]
D_{33}
\Bigg\}
[M_{h_0}^2, M_{h_0}^2,
0, M_{A_0}^2;
k_{12}, k_{23};
H^\pm, W, W, W]
\nonumber\\
&&
+
\dfrac{1}{M_W^2}
\Bigg\{
\Big[
M_{h_0}^2
(2 M_{H^\pm}^2+M_W^2)
-
2 M_{H^\pm}^4
+
M_W^2 (k_{12} - M_W^2)
\Big]
M_{h_0}^2
\nonumber \\
&&
-
M_{H^\pm}^2 M_W^2
(5 M_{H^\pm}^2-5 M_W^2-k_{12})
+
M_{A_0}^2
\Big[
M_{h_0}^2
(-2 M_{h_0}^2
+2 M_{H^\pm}^2
+M_W^2)
\nonumber \\
&&
+
M_W^2
(3 M_{H^\pm}^2
+2 M_W^2
-2 k_{12})
\Big]
\Bigg\}
D_0
[M_{h_0}^2, M_{h_0}^2,
0, M_{A_0}^2;
k_{12}, k_{23};
H^\pm, W, W, W]
\nonumber \\
&&
+
\dfrac{1}{M_W^2}
\Bigg\{
2 M_W^2 M_{A_0}^4
-
2 M_{A_0}^2
\Big[
M_W^2 (3 M_W^2+k_{12}+k_{13})
-
M_{h_0}^2
(-M_{h_0}^2
+M_{H^\pm}^2
+2 M_W^2)
\Big]
\nonumber \\
&&
+
2 M_{h_0}^4 M_{H^\pm}^2
+
M_W^2
\Big[
-3 M_{H^\pm}^4
+3 (M_W^2
+k_{12}
+k_{13})
M_{H^\pm}^2
+2 M_W^2
(k_{12}+k_{13})
\Big]
\nonumber \\
&&
-
M_{h_0}^2
\Big[
2 M_{H^\pm}^4
+3 M_W^2 M_{H^\pm}^2
+M_W^2
(M_W^2+k_{12}+k_{13})
\Big]
\Bigg\}
\times
\nonumber\\
&&
\times
D_3[M_{h_0}^2, M_{h_0}^2,
0, M_{A_0}^2;
k_{12}, k_{23};
H^\pm, W, W, W]
\nonumber \\
&&
+
\dfrac{1}{M_W^2}
\Bigg\{
-
M_{h_0}^2
\Big[
2 M_{H^\pm}^4+4 M_W^2 M_{H^\pm}^2
+M_W^2 (3 M_W^2+3 k_{12}+4 k_{13})
\Big]
\nonumber \\
&&
+
M_{h_0}^4
(2 M_{H^\pm}^2+3 M_W^2)
-
2 M_{A_0}^2
\Big[
M_{h_0}^4
-(M_{H^\pm}^2+2 M_W^2)
M_{h_0}^2+2 M_W^4
\Big]
\nonumber \\
&&
+M_W^2
\Big[
-3 M_{H^\pm}^4
+(3 M_W^2
+3 k_{12}
+4 k_{13})
M_{H^\pm}^2
+4 M_W^2 k_{13}
\Big]
\Bigg\}
\times
\nonumber\\
&&
\times
D_2[M_{h_0}^2, M_{h_0}^2,
0, M_{A_0}^2;
k_{12}, k_{23};
H^\pm, W, W, W]
\nonumber
\end{eqnarray}
\begin{eqnarray}
&&
+
\dfrac{1}{M_W^2}
\Bigg\{
2 M_{h_0}^4
(M_{H^\pm}^2+4 M_W^2)
\nonumber\\
&&
-
\Big[
(2 M_{H^\pm}^2
+
5 M_W^2) M_{H^\pm}^2
+
M_W^2
(5 M_W^2
+3 k_{12}
+ 3 k_{13})
\Big]
M_{h_0}^2
\nonumber \\
&&
+
M_{A_0}^2
\Big[
(-2 M_{h_0}^2+2 M_{H^\pm}^2-M_W^2)
M_{h_0}^2
+
M_W^2 (M_{H^\pm}^2-2 M_W^2)
\Big]
\nonumber \\
&&
+
M_W^2
\Big[
3 M_{H^\pm}^2
(- M_{H^\pm}^2+M_W^2
+k_{12}+k_{13})
+
2 M_W^2 (k_{12}+k_{13})
\Big]
\Bigg\}
\times
\nonumber\\
&&
\times
D_1
[M_{h_0}^2, M_{h_0}^2, 0, M_{A_0}^2;
k_{12}, k_{23};
H^\pm, W, W, W],
\end{eqnarray}
\begin{eqnarray}
\dfrac{G_4}
{M_W^2}
&=&
\Bigg\{
2 (M_{A_0}^2
-4 M_{h_0}^2
+3 M_{H^\pm}^2-3 M_W^2)
D_{00}
+
2 (M_{A_0}^2
-M_{H^\pm}^2-M_W^2)
\big(
M_{A_0}^2 D_{2}
+
k_{12} D_{3}
\big)
\nonumber \\
&&
+
2 (M_{A_0}^2-M_{H^\pm}^2-M_W^2)
(2 M_{A_0}^2
+M_{h_0}^2
+M_{H^\pm}^2-M_W^2-k_{12}-k_{13})
D_{0}
\nonumber \\
&&
+
\Big[
(-M_{A_0}^2
-6 M_{h_0}^2
+5 M_{H^\pm}^2
-8 M_W^2+k_{12}+2 k_{13})
M_{A_0}^2
\nonumber \\
&&
+
M_{H^\pm}^2 k_{12}
+
2 M_W^2 k_{13}
-
2 M_{h_0}^2
(M_W^2-M_{h_0}^2+k_{12}+k_{13})
\Big]
D_{11}
\nonumber \\
&&
+
\Big[
(2 M_{h_0}^2+2 M_{H^\pm}^2
-7 M_W^2-2 k_{12}+2 k_{13}) 
M_{h_0}^2+M_{H^\pm}^2 k_{12}
\nonumber \\
&&
+M_{A_0}^2 (-6 M_{h_0}^2
+2 M_{H^\pm}^2-3 M_W^2+k_{12})
+
(3 M_W^2
-
2 M_{H^\pm}^2) k_{13}
\Big]
D_{13}
\nonumber \\
&&
+
\Big[
(-M_{A_0}^2-10 M_{h_0}^2+5 M_{H^\pm}^2
-8 M_W^2+ 2 k_{12}+ 2 k_{13}) M_{A_0}^2
\nonumber \\
&&
+
M_{h_0}^2 (4 M_{h_0}^2
+2 M_{H^\pm}^2-9 M_W^2)
\nonumber \\
&&
-
(2 M_{H^\pm}^2-5 M_W^2) 
(k_{12}+k_{13})
\Big]
D_{12}
\Bigg\}
[M_{A_0}^2, M_{h_0}^2, 0, M_{h_0}^2;
k_{23}, k_{13};
H^\pm, W, W, W]
\nonumber \\
&&
+
\dfrac{1}{M_W^2}
\Bigg\{
\Big[
(-2 M_{h_0}^2+2 M_{H^\pm}^2
-3 M_W^2) M_{h_0}^2
+
M_W^2 (2 M_{A_0}^2
+3 M_{H^\pm}^2-k_{12}+k_{13})
\Big] M_{A_0}^2
\nonumber \\
&&
+
\Big[
M_W^2 (M_W^2-2 k_{12}-4 k_{13})
-
M_{H^\pm}^2
(M_{H^\pm}^2+k_{12}+3 k_{13})
\Big]
M_W^2
\nonumber \\
&&
+
\Big[
2 M_{h_0}^2 (M_{H^\pm}^2-M_W^2)
+
2 M_W^2 (3 M_W^2+k_{12}+k_{13})
+
M_{H^\pm}^2
(M_W^2 - 2 M_{H^\pm}^2)
\Big]
M_{h_0}^2
\Bigg\}
\times
\nonumber\\
&&
\times
D_1
[M_{A_0}^2, M_{h_0}^2, 0, M_{h_0}^2;
k_{23}, k_{13};
H^\pm, W, W, W],
\end{eqnarray}
\begin{eqnarray}
\dfrac{G_5}
{M_W^2}
&=&
\Bigg\{
4 (M_{A_0}^2-M_{h_0}^2-M_W^2)
D_{00}
-
\Big[
(M_{H^\pm}^2
-5 M_{h_0}^2+4 M_W^2+k_{12}
-k_{13}) M_{h_0}^2
\nonumber \\
&&
+
(M_{H^\pm}^2
+
2 M_W^2) k_{12}
+
M_{A_0}^2
(5 M_{h_0}^2
-M_{H^\pm}^2
-2 k_{12})
+M_{H^\pm}^2 k_{13}
\Big]
D_{1}
\nonumber \\
&&
-
\Big[
2 k_{12} (M_{h_0}^2+2 M_W^2)
+
M_{A_0}^2
(M_{h_0}^2
-M_{H^\pm}^2
+2 M_W^2-2 k_{12})
\Big]
D_{22}
\nonumber \\
&&
+
\Big[
(M_{h_0}^2
+3 M_{H^\pm}^2
-6 M_W^2-k_{12}
+k_{13}) M_{h_0}^2
-M_{H^\pm}^2 k_{12}
\nonumber \\
&&
+M_{A_0}^2 (-5 M_{h_0}^2
+M_{H^\pm}^2-2 M_W^2+2 k_{12})
+
(2 M_W^2 - M_{H^\pm}^2) k_{13}
\Big]
\big(
D_{12}
+
D_{13}
\big)
\nonumber \\
&&
-
\Big[
(2 M_W^2-M_{h_0}^2-M_{H^\pm}^2
+3 k_{12}+k_{13}) M_{h_0}^2
+
(M_{H^\pm}^2 + 4 M_W^2) k_{12}
\nonumber \\
&&
+
(M_{H^\pm}^2 - 2 M_W^2) k_{13}
+
M_{A_0}^2
(5 M_{h_0}^2
-3 M_{H^\pm}^2
+8 M_W^2
-4 k_{12}-2 k_{13})
\Big]
D_{23}
\nonumber \\
&&
-
\Big[
(-M_{h_0}^2+4 M_{A_0}^2
-M_{H^\pm}^2+2 M_W^2+k_{12}+k_{13})
M_{h_0}^2
+
M_{H^\pm}^2 k_{12}
+
(M_{H^\pm}^2 - 2 M_W^2) k_{13}
\nonumber \\
&&
-
2 M_{A_0}^2
(M_{H^\pm}^2
-3 M_W^2
+k_{12}
+k_{13})
\Big]
D_{33}
\Bigg\}
[M_{h_0}^2, M_{h_0}^2,
0, M_{A_0}^2;
k_{12}, k_{13};
H^\pm, W, W, W]
\nonumber \\
&&
+
\dfrac{1}{M_W^2}
\Bigg\{
M_{h_0}^4
(2 M_{H^\pm}^2+M_W^2)
-
\Big[
2 M_{H^\pm}^4
+M_W^2
(M_W^2+k_{12})
\Big]
M_{h_0}^2
\nonumber \\
&&
+M_W^2
\Big[
M_{H^\pm}^2
(-5 M_{H^\pm}^2
+
5 M_W^2
+
3 k_{12})
+
2 M_W^2 k_{12}
\Big]
\nonumber \\
&&
+
M_{A_0}^2
\Big[
(-2 M_{h_0}^2
+2 M_{H^\pm}^2
+3 M_W^2) M_{h_0}^2
+
M_W^2
(M_{H^\pm}^2
-2 k_{12})
\Big]
\Bigg\}
\times
\nonumber\\
&&
\times
D_0
[M_{h_0}^2, M_{h_0}^2, 0, M_{A_0}^2;
k_{12}, k_{13};
H^\pm, W, W, W]
\nonumber \\
&&
+
\dfrac{1}{M_W^2}
\Bigg\{
2 M_{h_0}^4 (M_{H^\pm}^2-M_W^2)
\nonumber \\
&&
+
\Big[
M_{H^\pm}^2
(3 M_W^2
-
2 M_{H^\pm}^2)
+
M_W^2 (3 M_W^2+2 k_{12}+k_{13})
\Big] M_{h_0}^2
\nonumber \\
&&
+M_W^2
\Big[
(-3 M_{H^\pm}^2
+3 M_W^2
-2 k_{12}
-3 k_{13})
M_{H^\pm}^2
-
2 M_W^2 (k_{12}+k_{13})
\Big]
\nonumber \\
&&
+
M_{A_0}^2
\Big[
(-2 M_{h_0}^2
+2 M_{H^\pm}^2
-3 M_W^2) M_{h_0}^2
+
M_W^2
(5 M_{H^\pm}^2-2 M_W^2+2 k_{13})
\Big]
\Bigg\}
\times
\nonumber\\
&&
\times
D_3
[M_{h_0}^2, M_{h_0}^2,
0, M_{A_0}^2;
k_{12}, k_{13};
H^\pm, W, W, W]
\nonumber \\
&&
+
\dfrac{1}{M_W^2}
\Bigg\{
\Big[
M_{h_0}^2
(2 M_{H^\pm}^2-5 M_W^2)
+
2 M_{H^\pm}^2
(2 M_W^2
- M_{H^\pm}^2)
\nonumber \\
&&
+
M_W^2 (5 M_W^2+3 k_{12}+4 k_{13})
\Big]
M_{h_0}^2
\nonumber \\
&&
-
2 M_{A_0}^2
\Big[
(M_{h_0}^2+M_W^2-M_{H^\pm}^2) M_{h_0}^2
-
M_W^2 (3 M_{H^\pm}^2+M_W^2)
\Big]
\nonumber \\
&&
+
M_W^2
\Big[
M_{H^\pm}^2
(-3 M_{H^\pm}^2
+3 M_W^2
-3 k_{12}
-4 k_{13})
-
2 M_W^2
(3 k_{12} + 2 k_{13})
\Big]
\Bigg\}
\times
\nonumber\\
&&
\times
D_2
[M_{h_0}^2, M_{h_0}^2, 0, M_{A_0}^2;
k_{12}, k_{13};
H^\pm, W, W, W].
\end{eqnarray}
We next to decompose
the one-loop form factors
$F^{(\textrm{Box})}_{(SS,W)}$
as follows:
\begin{eqnarray}
\dfrac{
F^{(\textrm{Box})}_{
(SS,W)}
}{k_1\cdot k_3}
&=&
-
\frac{e^2}{8\pi^2}
\left(
\frac{
g_{h_0 h_0 H^\pm G^\mp}
}
{M_W s_W}
\right)
\cdot 
H_0
\\
&&
-
\frac{ie^3}{16\pi^2}
\left(
\frac{
c_{\beta-\alpha}
}{
M_W^2
s_{2W}
}
\right)
g_{h_0 H^\pm H^\mp}
\cdot
\sum
\limits_{i = 1}^{3}
H_i
\nonumber \\
&&
-
\dfrac{e^4}{16\pi^2}
\left(
\dfrac{
s_{2(\beta-\alpha)}}
{M_W^3s_{2W}^2\; c_W}
\right)
\cdot 
\sum
\limits_{i = 4}^{6}
H_i.
\nonumber
\end{eqnarray}
Both general couplings
in the above equation 
are obtained as in Eqs.~\eqref{hHH}
and \eqref{hhHG}. All coefficient 
factors $H_i$ 
for $i=0,1,\cdots, 6$ 
given in the above equation,
are listed in terms of
PV-functions as follows:
\begin{eqnarray}
\dfrac{H_0}{M_W^2}
&=&
\dfrac{
\big(
M_{A_0}^2-M_{H^\pm}^2
\big)
}{M_W^2}
C_2[0, k_{12}, M_{A_0}^2;
H^\pm, H^\pm, W],
\\
&& 
\nonumber
\\
\dfrac{H_1}{M_W^2}
&=&
\Big[
C_2
-
C_1
-
C_0
\Big]
[M_{h_0}^2, k_{12}, M_{h_0}^2;
H^\pm, H^\pm, W]
\nonumber
\\
&&
+
\Big[
C_2
+
C_1
-
C_0
\Big]
[M_{h_0}^2, k_{23}, M_{A_0}^2;
H^\pm, H^\pm, W]
-
2 C_0
[M_{h_0}^2, 0, k_{23}; H^\pm, W, W]
\nonumber \\
&&
+
\Big[
C_2
-
C_0
\Big]
[M_{h_0}^2, k_{13}, M_{A_0}^2;
H^\pm, H^\pm, W]
-
4 C_0
[M_{h_0}^2, 0, k_{13}; H^\pm, W, W],
\\
&& 
\nonumber
\\
\dfrac{H_2}{M_W^2
}
&=&
\Bigg[
4 D_{00}
+
(M_{A_0}^2-M_{h_0}^2+k_{12}-k_{13})
\big( D_{12} + D_{13} \big)
+
(M_{A_0}^2-5 M_{h_0}^2+k_{12}-k_{13})
D_1
\nonumber \\
&&
+
(3 M_{A_0}^2-M_{h_0}^2+3 k_{12}+k_{13})
D_{23}
+
(2 M_{A_0}^2-M_{h_0}^2+k_{12}+k_{13})
D_{33}
\nonumber \\
&&
+
(M_{A_0}^2+2 k_{12})
D_{22}
\Bigg]
[M_{h_0}^2, M_{h_0}^2, 0, M_{A_0}^2;
k_{12}, k_{13};
H^\pm, H^\pm, W, W]
\nonumber \\
&&
+
\dfrac{1}{M_W^2}
\Bigg[
M_W^2
(-M_{A_0}^2-M_{h_0}^2
-2 M_{H^\pm}^2+k_{12})
+
2 (M_{A_0}^2 - M_{H^\pm}^2)
(M_{h_0}^2 - M_{H^\pm}^2)
\Bigg]
\times
\nonumber\\
&&
\times
D_0
[M_{h_0}^2, M_{h_0}^2, 0, M_{A_0}^2;
k_{12}, k_{13};
H^\pm, H^\pm, W, W]
\nonumber \\
&&
+
\Big[
M_W^2
(4 M_{A_0}^2+5 M_{h_0}^2
-2 M_{H^\pm}^2-3 k_{12}
-4 k_{13})
+
2 (M_{A_0}^2 - M_{H^\pm}^2)
(M_{h_0}^2 - M_{H^\pm}^2)
\Big]
\times
\nonumber\\
&&
\times
D_2
[M_{h_0}^2, M_{h_0}^2, 0, M_{A_0}^2;
k_{12}, k_{13};
H^\pm, H^\pm, W, W]
\nonumber \\
&&
+
\dfrac{1}{M_W^2}
\Bigg[
M_W^2
(3 M_{A_0}^2+2 M_{h_0}^2
-2 M_{H^\pm}^2 -2 k_{12} -k_{13})
+
2 (M_{A_0}^2 - M_{H^\pm}^2)
(M_{h_0}^2 - M_{H^\pm}^2)
\Bigg]
\times
\nonumber\\
&&
\times
D_3
[M_{h_0}^2, M_{h_0}^2, 0, M_{A_0}^2;
k_{12}, k_{13};
H^\pm, H^\pm, W, W]
.
\end{eqnarray}
The further coefficient 
factors are given by
\begin{eqnarray}
\dfrac{H_3}{M_W^2}
&=&
\Bigg[
2 D_{00}
-
(3 M_{h_0}^2-2 k_{12}-k_{13})
D_{11}
+
(M_{A_0}^2-3 M_{h_0}^2
+4 k_{12}+k_{13})
D_{12}
\nonumber \\
&&
\hspace{0cm}
+
(3 M_{A_0}^2-2 M_{h_0}^2+2 k_{12})
D_{13}
+
(3 M_{A_0}^2+M_{h_0}^2-k_{13})
D_{33}
\nonumber \\
&&\hspace{0cm}
+
(4 M_{A_0}^2+M_{h_0}^2+2 k_{12}-k_{13})
D_{23}
\nonumber \\
&&\hspace{0cm}
+
(M_{A_0}^2+2 k_{12})
D_{22}
\Bigg]
[M_{h_0}^2, M_{h_0}^2, 0, M_{A_0}^2;
k_{12}, k_{23};
H^\pm, H^\pm, W, W]
\nonumber \\
&&
+
\dfrac{1}{M_W^2}
\Bigg[
M_{A_0}^2 (2 M_{h_0}^2
-2 M_{H^\pm}^2+M_W^2)
-M_{h_0}^2 (2 M_{H^\pm}^2+M_W^2)
\nonumber\\
&&
+
2 M_{H^\pm}^2
(M_{H^\pm}^2 - M_W^2)
-
M_W^2 k_{12}
\Bigg]
D_0
[M_{h_0}^2, M_{h_0}^2
, 0, M_{A_0}^2;
k_{12}, k_{23};
H^\pm, H^\pm, W, W]
\nonumber \\
&&
+
\dfrac{1}{M_W^2}
\Bigg[
M_W^2
(-M_{A_0}^2-8 M_{h_0}^2
-2 M_{H^\pm}^2
+3 k_{12}+3 k_{13})
\nonumber\\
&&
+
2 (M_{A_0}^2 - M_{H^\pm}^2)
(M_{h_0}^2 - M_{H^\pm}^2)
\Bigg]
D_1[M_{h_0}^2, M_{h_0}^2
, 0, M_{A_0}^2;
k_{12}, k_{23};
H^\pm, H^\pm, W, W]
\nonumber \\
&&
+
\dfrac{1}{M_W^2}
\Bigg[
M_W^2
(-2 M_{A_0}^2 -2 M_{H^\pm}^2
-3 M_{h_0}^2+3 k_{12}+4 k_{13})
\nonumber\\
&&
+
2 (M_{A_0}^2 - M_{H^\pm}^2)
(M_{h_0}^2 - M_{H^\pm}^2)
\Bigg]
D_2
[M_{h_0}^2, M_{h_0}^2
, 0, M_{A_0}^2;
k_{12}, k_{23};
H^\pm, H^\pm, W, W]
\nonumber \\
&&
+
\Big[
2 (M_{A_0}^2 - M_{H^\pm}^2)
(M_{h_0}^2 - M_{H^\pm}^2)
-
M_W^2
(2 M_{H^\pm}^2-k_{12}-k_{13})
\Bigg]
\times
\nonumber\\
&&
\times
D_3
[M_{h_0}^2, M_{h_0}^2,
0, M_{A_0}^2;
k_{12}, k_{23};
H^\pm, H^\pm, W, W],
\\
&&
\nonumber
\\
\dfrac{H_4}{M_W^2}
&=&
\dfrac{1}{M_W^2}
\Bigg[
M_{A_0}^2
(M_{h_0}^2-M_{H^\pm}^2-M_W^2)
-
2 M_W^4
-
M_{h_0}^2 (M_{H^\pm}^2+M_W^2)
\nonumber\\
&&
\hspace{1cm}
+
M_{H^\pm}^2
(M_{H^\pm}^2 + 2 M_W^2)
\Bigg]
C_2[0, k_{13}, M_{h_0}^2;
H^\pm, H^\pm, W],
\end{eqnarray}
\begin{eqnarray}
\dfrac{H_5}{M_W^2}
&=&
\Bigg\{
(2 M_{A_0}^2
+ 2 M_{h_0}^2
- 4 M_{H^\pm}^2+ 8 M_W^2)
D_{00}
-
\Big[
M_{A_0}^2 (M_{h_0}^2
-2 M_{H^\pm}^2+3 M_W^2+k_{13})
\nonumber \\
&&\hspace{0cm}
+
M_{h_0}^2
(M_{h_0}^2-M_W^2-2 k_{12}-k_{13})
+
(2 M_{H^\pm}^2 - 3 M_W^2)
k_{12}
+
M_W^2 k_{13}
\Big]
\big(
D_{12}
+
D_{13}
\big)
\nonumber \\
&&\hspace{0cm}
+
\Big[
M_{A_0}^2 (M_{h_0}^2-M_{H^\pm}^2+2 M_W^2+k_{12})
+
k_{12} (2 M_{h_0}^2-3 M_{H^\pm}^2+5 M_W^2)
\Big]
D_{22}
\nonumber \\
&&\hspace{0cm}
+
\Big[
M_{A_0}^2 (3 M_{h_0}^2-M_{H^\pm}^2
+2 M_W^2+k_{12})
+
M_{h_0}^2 (-3 M_{h_0}^2 + M_{H^\pm}^2
+4 k_{12}+k_{13})
\nonumber \\
&&\hspace{0cm}
+
(8 M_W^2
-
5 M_{H^\pm}^2) k_{12}
+
(2 M_W^2
-
M_{H^\pm}^2) k_{13}
\Big]
D_{23}
\nonumber \\
&&\hspace{0cm}
+
\Big[
M_{h_0}^2
(2 M_{A_0}^2-3 M_{h_0}^2
+M_{H^\pm}^2+2 k_{12}+k_{13})
+
(3 M_W^2
-
2 M_{H^\pm}^2) k_{12}
\nonumber \\
&&\hspace{0cm}
+
(2 M_W^2
-
M_{H^\pm}^2) k_{13}
\Big]
D_{33}
\Bigg\}
[0, M_{A_0}^2, M_{h_0}^2, M_{h_0}^2;
k_{12}, k_{13};
H^\pm, H^\pm, W, W]
\nonumber\\
&&
+
\dfrac{1}{M_W^2}
\Bigg\{
M_{A_0}^2
\Big[
M_{h_0}^2
(M_{h_0}^2 - M_{H^\pm}^2 - 2 M_W^2)
+
M_W^2 (M_{H^\pm}^2-M_W^2+2 k_{12})
\Big]
\nonumber\\
&&
\hspace{0cm}
+
M_{h_0}^2
\Big[
M_{H^\pm}^2
(M_{H^\pm}^2 + 3 M_W^2)
-
M_{h_0}^2
(M_{H^\pm}^2+2 M_W^2)
+
M_W^2
(M_W^2 +2 k_{12} + 2 k_{13})
\Big]
\nonumber\\
&&
\hspace{0cm}
+ M_W^2
\Big[
2 M_{H^\pm}^2
(M_W^2-2 k_{12}-k_{13})
+
M_W^2
(-2 M_W^2+k_{12}-2 k_{13})
\Big]
\Bigg\}
\times
\nonumber\\
&&
\times
\big(D_2+D_3\big)
[0, M_{A_0}^2, M_{h_0}^2, M_{h_0}^2;
k_{12}, k_{13};
H^\pm, H^\pm, W, W],
\end{eqnarray}
\begin{eqnarray}
\dfrac{H_6}{M_W^2}
&=&
\Bigg
\{
(4 M_{h_0}^2-4 M_{H^\pm}^2+6 M_W^2)
D_{00}
-
\Big[
M_{A_0}^2
(M_{A_0}^2+2 M_{h_0}^2
-2 M_{H^\pm}^2+4 M_W^2-k_{12}-k_{13})
\nonumber \\
&&
+
M_{h_0}^2
(-M_{h_0}^2+M_W^2-k_{12}+k_{13})
+
(2 M_{H^\pm}^2-4 M_W^2)
k_{12}
-
M_W^2 k_{13}
\Big]
D_{12}
\nonumber \\
&&
+
\Big[
M_{A_0}^2
(M_{h_0}^2-M_{H^\pm}^2+2 M_W^2+k_{12})
+
k_{12}
(2 M_{h_0}^2-3 M_{H^\pm}^2+5 M_W^2)
\Big]
D_{22}
\nonumber \\
&&
+
\Big[
M_{A_0}^2 (3 M_{h_0}^2
-M_{H^\pm}^2+2 M_W^2)
\nonumber \\
&&
-
M_{h_0}^2
(M_{h_0}^2+M_{H^\pm}^2
-4 M_W^2-k_{12}+k_{13})
-
M_W^2 k_{13}
\nonumber \\
&&
+
(M_{H^\pm}^2 - M_W^2)
(k_{13} - k_{12})
\Big]
D_{23}
\Bigg\}
[0, M_{A_0}^2, M_{h_0}^2, M_{h_0}^2;
k_{12}, k_{23};
H^\pm, H^\pm, W, W]
\nonumber\\
&&
+
\dfrac{1}{M_W^2}
\Bigg\{
M_{A_0}^2
\Big[
M_{h_0}^2
(M_{h_0}^2 - M_{H^\pm}^2)
-
M_W^2
(M_{H^\pm}^2+3 M_W^2-2 k_{12})
\Big]
\nonumber \\
&&
+
M_{h_0}^2
\Big[
M_{H^\pm}^2
(M_{H^\pm}^2 - M_W^2)
-
M_{h_0}^2
(M_{H^\pm}^2-2 M_W^2)
-
M_W^2
(3 M_W^2+2 k_{13})
\Big]
\nonumber \\
&&
+
M_W^2
\Big[
2 M_{H^\pm}^2
(M_W^2-k_{12}+k_{13})
-
M_W^2
(2 M_W^2-3 k_{12}-2 k_{13})
\Big]
\Bigg\}
\times
\nonumber \\
&&
\times
D_2
[0, M_{A_0}^2, M_{h_0}^2, M_{h_0}^2;
k_{12}, k_{23};
H^\pm, H^\pm, W, W].
\end{eqnarray}
Finally, we consider
the last form factors
$F^{(\textrm{Box})}_{
(SSS, W)}$
which are shown as follows:
\begin{eqnarray}
\dfrac{
F^{(\textrm{Box})}_{
(SSS, W)
}
}
{k_1\cdot k_3}
&=&
\frac{ie^3}{16\pi^2}
\left(
\frac{c_{\beta-\alpha}}{M_W^2s_W^2}
\right)
\cdot 
\sum
\limits_{i = 0}^{5}
K_i.
\end{eqnarray}
In this equation,
all factors $K_i$
are listed in terms
of PV-functions
as follows:
\begin{eqnarray}
\dfrac{K_0}
{M_W^2}
&=&
-
\big(
C_0
+
C_1
+
C_2
\big)
[M_{h_0}^2,0,k_{13};
H^\pm, H^\pm, H^\pm],
\\
&&
\nonumber\\
\dfrac{K_1}
{M_W^2
}
&=&
-
\Bigg[
2 D_{00}
-
(M_{A_0}^2+3 M_{h_0}^2-k_{13})
D_{23}
-
(M_{A_0}^2+3 M_{h_0}^2-k_{13})
D_{13}
\nonumber \\
&&
-
(3 M_{h_0}^2-2 k_{12}-k_{13})
D_{11}
+
(2 M_{A_0}^2-5 M_{h_0}^2+2 k_{12}+k_{13})
D_{12}
+
2 (M_{A_0}^2 - M_{h_0}^2)
D_{22}
\nonumber \\
&&
-
(M_{A_0}^2+3 M_{h_0}^2-k_{13})
D_3
\Bigg]
[M_{h_0}^2,0,M_{A_0}^2,M_{h_0}^2;
k_{13},k_{12};
H^\pm, H^\pm, H^\pm, W]
\nonumber
\\
&&
+
\Bigg[
M_{A_0}^2
(M_{h_0}^2-M_{H^\pm}^2)
-
M_W^2
(M_{A_0}^2-M_{h_0}^2+M_W^2-k_{13})
+
M_{H^\pm}^2
(M_{H^\pm}^2-M_{h_0}^2)
\Bigg]\times
\nonumber\\
&&
\times
D_0
[M_{h_0}^2,0,M_{A_0}^2,M_{h_0}^2;
k_{13},k_{12};
H^\pm, H^\pm, H^\pm, W]
\nonumber
\\
&&
+
\Bigg[
M_{A_0}^2
(M_{h_0}^2-M_{H^\pm}^2)
-
M_W^2
(M_{A_0}^2-4 M_{h_0}^2+
M_W^2+2 k_{12})
+
M_{H^\pm}^2
(M_{H^\pm}^2-M_{h_0}^2)
\Bigg]
\times
\nonumber\\
&&
\times
D_1
[M_{h_0}^2,0,M_{A_0}^2,M_{h_0}^2;
k_{13},k_{12};
H^\pm, H^\pm, H^\pm, W]
\nonumber \\
&&
+
\Bigg[
M_{A_0}^2
(M_{h_0}^2-M_{H^\pm}^2)
-
M_W^2
(3 M_{A_0}^2-3 M_{h_0}^2+
M_W^2-k_{13})
+
M_{H^\pm}^2
(M_{H^\pm}^2-M_{h_0}^2)
\Bigg]
\times
\nonumber\\
&&
\times
D_2
[M_{h_0}^2,0,M_{A_0}^2,M_{h_0}^2;
k_{13},k_{12};
H^\pm, H^\pm, H^\pm, W],
\\
&& \nonumber\\
\dfrac{K_2}{M_W^2}
&=&
\Bigg[
-4 D_{00}
+
(M_{A_0}^2-M_{h_0}^2+k_{12}-k_{13})
D_{13}
-
(M_{h_0}^2+k_{12}+k_{13})
D_{33}
\nonumber \\
&&
+
2 (M_{A_0}^2 - M_{h_0}^2)
D_{23}
\Bigg]
[M_{h_0}^2,0,M_{A_0}^2,
M_{h_0}^2;k_{23},k_{12};
H^\pm, H^\pm, H^\pm, W],
\end{eqnarray}
\begin{eqnarray}
\dfrac{K_3}{M_W^2}
&=&
-
\Bigg[
4 D_{00}
-
2 (M_{A_0}^2 - M_{h_0}^2)
D_{23}
+
(M_{h_0}^2+k_{12}+k_{13})
D_{33}
\nonumber\\
&&
-
(M_{A_0}^2-M_{h_0}^2-k_{12}+k_{13})
D_{13}
\Bigg]
[0,M_{h_0}^2,M_{A_0}^2,M_{h_0}^2;
k_{23},k_{13};
H^\pm, H^\pm, H^\pm, W]
\nonumber \\
&&
-
\dfrac{1}{M_W^2}
\Bigg[
M_{A_0}^2
(M_{h_0}^2-M_{H^\pm}^2)
-
M_W^2
(k_{12}+k_{13}-3M_{h_0}^2+M_W^2)
+
M_{H^\pm}^2
(M_{H^\pm}^2-M_{h_0}^2)
\Bigg]
\times
\nonumber\\
&&
\times
D_3
[0,M_{h_0}^2,M_{A_0}^2,M_{h_0}^2;
k_{23},k_{13};
H^\pm, H^\pm, H^\pm, W],
\end{eqnarray}
\begin{eqnarray}
\dfrac{K_4}{M_W^2}
&=&
-\Bigg[
(k_{12} - 2 M_{h_0}^2)
D_{11}
+
(M_{A_0}^2-3 M_{h_0}^2+k_{12}-k_{13})
D_{12}
+
(M_{A_0}^2+M_{h_0}^2-k_{13})
D_{33}
\nonumber \\
&&
+
(M_{A_0}^2-M_{h_0}^2+k_{12}-k_{13})
D_{13}
+
(M_{A_0}^2-M_{h_0}^2-k_{13})
D_{22}
\nonumber \\
&&
+
2 (M_{A_0}^2 - k_{13})
D_{23}
\Bigg]
[M_{h_0}^2,0,M_{h_0}^2,
M_{A_0}^2;k_{13},k_{23};
H^\pm, H^\pm, H^\pm, W]
\nonumber \\
&&
-
\dfrac{1}{M_W^2}
\Bigg[
M_{A_0}^2
(M_{h_0}^2-M_{H^\pm}^2)
-
M_W^2
(M_{h_0}^2+M_{H^\pm}^2-k_{13})
+
M_{H^\pm}^2
(M_{H^\pm}^2-M_{h_0}^2)
\Bigg]
\times
\nonumber\\
&&
\times
D_0
[M_{h_0}^2,0,M_{h_0}^2,
M_{A_0}^2;k_{13},k_{23};
H^\pm, H^\pm, H^\pm, W]
\nonumber \\
&&
-
\dfrac{1}{M_W^2}
\Bigg[
M_{A_0}^2
(M_{h_0}^2-M_{H^\pm}^2)
-
M_W^2
(3 M_{h_0}^2
+M_{H^\pm}^2
-k_{12}-k_{13})
+
M_{H^\pm}^2
(M_{H^\pm}^2-M_{h_0}^2)
\Bigg]
\times
\nonumber\\
&&
\times
D_1
[M_{h_0}^2,0,M_{h_0}^2,
M_{A_0}^2;k_{13},k_{23};
H^\pm, H^\pm, H^\pm, W]
\nonumber \\
&&
-
\dfrac{1}{M_W^2}
\Bigg[
M_W^2
(M_{A_0}^2
-2 M_{h_0}^2-M_{H^\pm}^2)
+
M_{A_0}^2
(M_{h_0}^2-M_{H^\pm}^2)
+
M_{H^\pm}^2
(M_{H^\pm}^2-M_{h_0}^2)
\Bigg]
\times
\nonumber\\
&&
\times
D_2
[M_{h_0}^2,0,M_{h_0}^2,
M_{A_0}^2;k_{13},k_{23};
H^\pm, H^\pm, H^\pm, W]
\nonumber \\
&&
-
\dfrac{1}{M_W^2}
\Bigg[
M_{A_0}^2
(M_{h_0}^2-M_{H^\pm}^2)
+
M_W^2
(M_{A_0}^2-M_{H^\pm}^2)
+
M_{H^\pm}^2
(M_{H^\pm}^2-M_{h_0}^2)
\Bigg]
\times
\nonumber\\
&&
\times
D_3
[M_{h_0}^2,0,M_{h_0}^2,
M_{A_0}^2;k_{13},k_{23};
H^\pm, H^\pm, H^\pm, W],
\end{eqnarray}
\begin{eqnarray}
\dfrac{K_5}{M_W^2}
&=&
-
\dfrac{1}{M_W^2}
\Bigg[
M_{A_0}^2
(M_{h_0}^2-M_{H^\pm}^2)
-
M_W^2
(k_{12}+k_{13}-3M_{h_0}^2+M_W^2)
+
M_{H^\pm}^2
(M_{H^\pm}^2-M_{h_0}^2)
\Bigg]
\times
\nonumber
\\
&&
\times
D_3[M_{h_0}^2,0,M_{A_0}^2,M_{h_0}^2
;k_{23},k_{12}; H^\pm, H^\pm, H^\pm, W].
\end{eqnarray}

After collecting all the necessary 
form factors, we are going to check 
for the analytic results with verifying
the $UV$-, $IR$-finiteness and Ward identity
of one-loop amplitude. The numerical tests
are shown in Tables~\ref{UV},~\ref{WardID}.
From the data, we find that the numerical 
results of the tests are good stabilities. 
Having the corresness one-loop form factors, 
the decay rates are then calculated
as follows:
\begin{eqnarray}
\Gamma_{A_0
\rightarrow h_0h_0\gamma}
&=&
\dfrac{1}{256 \pi^3 M_{A_0}^3}
\int \limits_{k_{12}^\text{min}}^{k_{12}^\text{max}}
d k_{12}
\int \limits_{k_{13}^\text{min}}^{k_{13}^\text{max}}
d k_{13}
\,
\sum \limits_\text{pol.}
|\mathcal{M}_{A_0
\rightarrow h_0h_0\gamma}|^2
\end{eqnarray}
where total amplitude is squared as
\begin{eqnarray}
\sum \limits_\text{unpol.}
|\mathcal{M}_{A_0
\rightarrow h_0h_0\gamma} |^2
&=&
\dfrac{
4[
k_{12} 
k_{13} 
k_{23}
-
M_{h_0}^2
(
M_{h_0}^2
-
M_{A_0}^2
)
k_{12}
-
M_{A_0}^4
M_{h_0}^2
]
}{
\big(
M_{h_0}^2-k_{13}
\big)^2 
\big(
M_{h_0}^2-k_{23}
\big)^2
}
\big|
F
\big|^2.
\end{eqnarray}
The integration regions are in
\begin{eqnarray}
k_{12}^\text{min}
&=& 4 M_{h_0}^2,
\quad 
k_{12}^\text{max}
=
M_{A_0}^2,
\\
k_{13}^\text{max/min}
&=&
\frac{1}{2}
\Big[
M_{A_0}^2
+
2 M_{h_0}^2
-
k_{12}
\pm
\big(
M_{A_0}^2
-
k_{12}
\big)
\sqrt{
1
-
4 M_{h_0}^2/k_{12}
}
\Big].
\end{eqnarray}
In the next section, we are going 
to present phenomenological results
for the decay process.
\section{Phenomenological results}
In phenomenological results, all
physical input parameters in the SM
are taken as same as in Ref.~\cite{Phan:2024zus}.
We are interested in the nearly
alignment scenario, or taking 
$s_{\beta-\alpha}
\rightarrow 1$ in this work.
In particular, 
$s_{\beta-\alpha}=0.95$ is 
selected for the following 
numerical analyses. While the mixing angle
$\beta$ is taken in the range of
$2\leq t_{\beta}< 8$.
Furthermore, the soft-breaking scale
for the $Z_2$-symmetry 
is obtained as $M^2=M_H^2$
in this work. For the decay widths of
$Z$ and the SM-like Higgs, we take
their values as in Ref.~\cite{Phan:2024zus}. 
In other cases, the decay widths 
of $A_0^*$ ($H^*$)
can be obtained as in Ref.~\cite{Aiko:2022gmz}
(in Ref.~\cite{Kanemura:2022ldq}), respectively.

In Table~\ref{TotalDecay}, the decay rates
for $A_0\rightarrow h_0h_0\gamma$ at 
several points in the parameter 
space of THDM are calculated. 
In the Table, we select 
$t_{\beta}=3$, $M_{A_0}=M_{H^\pm }$. 
While the mass of 
CP-even Higgs is taken as $M_H=M_{A_0}-M_Z$. 
In this Table, the first column shows for 
the values of $500$ GeV$\leq 
M_{A_0} \leq 1200$ GeV. The numerical 
results for the decay rates 
of $\Gamma_{A_0\rightarrow h_0h_0\gamma}$ 
are presented in the remaining columns which 
are corresponding to each type of THDM.
The results show that the decay rates 
are proportional to $M_{A_0}$. The decay 
widths for this mode are very small for 
the mass regions $M_{A_0}\leq 500$ GeV. They
are order of $\mathcal{O}(1)$ KeV 
in the regions of $M_{A_0}\geq 800$ GeV. We 
find that the results are lightly different 
from the distinct four types of the THDM.
It is because the distinct four types
of the THDM only come from all fermion exchanging in 
one-loop triangle diagrams with $Z^*$-poles
(seen Eqs.~\eqref{zfermion13},~\eqref{zfermion23} 
for more detail).
The contributions are proportional to the 
couplings $g_{h_0ff}$ and give 
small contributions in comparison 
with other terms. 
\begin{table}[H]
\centering
\begin{tabular}{|c|l|l|l|l|}
\hline\hline 
$M_{A_0}$
[GeV]
&
$
\Gamma^{\textrm{(I)}}_{A_0\rightarrow h_0h_0\gamma}
$~[KeV]
&
$
\Gamma^{\textrm{(II)}}_{A_0\rightarrow h_0h_0\gamma}
$~[KeV]
&
$
\Gamma^{\textrm{(X)}}_{A_0\rightarrow h_0h_0\gamma}
$~[KeV]
&
$
\Gamma^{\textrm{(Y)}}_{A_0\rightarrow h_0h_0\gamma}
$~[KeV]
\\
\hline
$500$
&
$0.008772$
&
$0.008793$
&
$0.008773$
&
$0.008789$
\\ 
&
$\pm 0.000009$
&
$\pm 0.000009$
&
$\pm 0.000009$
&
$\pm 0.000009$
\\
\hline
$800$
&
$0.9325 \pm 0.0009$
&
$0.9333 \pm 0.0009$
&
$0.9323 \pm 0.0009$
&
$0.9332 \pm 0.0009$
\\
\hline
$1000$
&
$5.378 \pm 0.005$
&
$5.380  \pm 0.005$
&
$5.378 \pm 0.005$
&
$5.380 \pm 0.005$
\\
\hline
$1200$
&
$19.11 \pm 0.02 $
&
$19.11 \pm 0.02 $
&
$19.11 \pm 0.02 $
&
$19.11 \pm 0.02 $
\\
\hline
\hline
\end{tabular}
\caption{
\label{TotalDecay} The decay rates for
the decay processs. In this Table, we chose 
$M_{H} = M_{A_0} - M_Z$, 
$s_{\beta - \alpha} = 0.95$ and 
$M_{H^\pm}= M_{A_0}$. Numerical results
for the partial decay rates 
are generated by using the multidimensional
numerical integration CUBA 
program~\cite{Hahn:2004fe}. }
\end{table}

\begin{figure}[H]
\centering
\hspace{-6.5cm}
$
d\Gamma_{A_0\rightarrow h_0h_0\gamma}/
dM_{h_0h_0}
\times 10^{-8}$
\\
\includegraphics[width=12cm, height=8cm]
{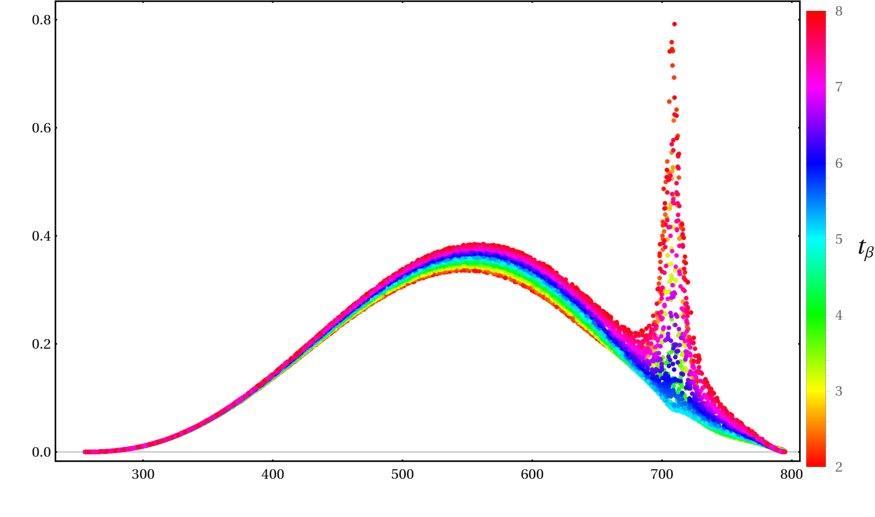}
\\ 
\hspace{10cm}$M_{h_0h_0}$~[GeV]
\caption{\label{dgammaHH}
The differential decay rates with respect to
the invariant of Higgs-pair ($M_{h_0h_0}$) 
in final state are scanned over
the parameter $t_\beta$. It is stress
that we have used the label
$
d\Gamma_{A_0\rightarrow h_0h_0\gamma}/
dM_{h_0h_0}
\times 10^{-8}$. 
It means that numerical results
shown in vertical axis must be multiplied by
the factor $10^{-8}$ in this label.
}
\end{figure}

In Fig.~\ref{dgammaHH}, the differential 
decay rates with respect to
the invariant of Higgs-pair ($M_{h_0h_0}$) 
in final state are scanned over
the parameter $t_\beta$.
We have already pointed out that 
the results are lightly different from 
considering the different types of 
the THDM. For this reason, we assume 
that there are no different results 
in the differential decay rates
from varying the obvious types 
of the THDM. In the below plots, 
we take THDM type I as typical example.
Additionally, we vary the mixing angle
$2\leq t_\beta \leq 8$.
We also set 
$M_{A_0}=800$ GeV, 
$M_{H} = M_{A_0} - M_Z$, 
$s_{\beta - \alpha} = 0.95$ and 
$M_{H^\pm}= M_{A_0}$ 
correspondingly. The decay rates 
develop to the regions of
$500$ GeV $\leq M_{h_0h_0}
\leq 600$ GeV and they are 
then decreased to $M_{h_0h_0}
\sim 680$ GeV. The decay rates
are increased rapidly
to the peak around 
$M_H \sim 710$ GeV. 
The peaks are suppressed 
in the case of $4 \leq t_{\beta}
\leq 6$, it may come from 
the cancellation of  one-loop 
triangle and the mixing 
$Z$-$\gamma$ with $H^*$-pole 
diagrams.

\begin{figure}[H]
\centering
\hspace{-6cm}
$
d\Gamma_{A_0\rightarrow h_0h_0\gamma}/
dM_{h_0h_0}
\times 10^{-8}$
\\
\includegraphics[width=12cm, height=8cm]
{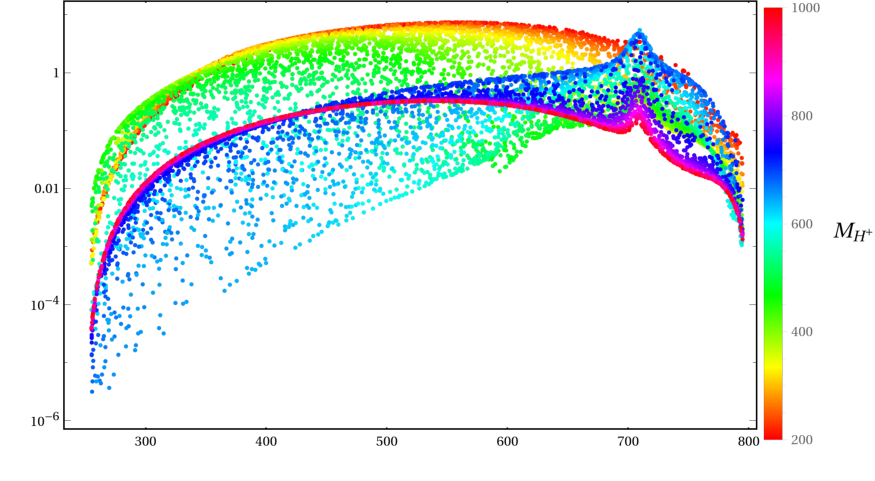}
\\ 
\hspace{10cm}$M_{h_0h_0}$~[GeV]
\caption{\label{dgammaHHMP}
The differential decay rates 
with respect to
the invariant of Higgs-pair 
($M_{h_0h_0}$) 
in final state are scanned over
$M_{H^{\pm}}^2$.
}
\end{figure}

In Fig.~\ref{dgammaHHMP}, 
the differential 
decay rates with respect to
the invariant of Higgs-pair ($M_{h_0h_0}$) 
in final state are scanned over the charged 
Higgs mass $M_{H^\pm}$
(take THDM type I as a typical example). 
The input parameters are used 
as in the previous plot and fix the mixing 
angle $t_{\beta}=3$. 
The results are generated with changing
$200$ GeV $\leq M_{H^\pm} \leq 1000$
GeV. 
Overall, the differential 
decay rates develop to 
the regions $\sim 350$ GeV
$\leq M_{h_0h_0} \leq \sim 730$ GeV. 
They change slightly in the mentioned 
regions and decrease rapidly when 
$M_{h_0h_0} \geq \sim 730$ GeV. 
We also find small peaks 
of the differential decay rates 
around $M_H\sim 710$ GeV. 
\section{Conclusions} 
For the first time, the analytic expressions for one-loop contributions to the rare decay amplitude  
$A_0 \rightarrow h_0h_0\gamma$ within the CP-conserving of the THDM have been introduced. They are written in term of scalar one-loop PV-functions in standard output of both the packages~{\tt LoopTools} and {\tt Collier}. Numerical checks for the calculations such as the $UV$-, $IR$-finiteness,  the Ward identity
of one-loop amplitude are also performed in this article. In phenomenological results,  the decay rates for $A_0 \rightarrow h_0h_0\gamma$  are examined at several allowed points in the parameter space of the THDM. Furthermore, the differential decay widths with respect to the invariant mass of Higgs-pair in final states are studied in the parameter space of the THDM. 
\\

\noindent
{\bf Acknowledgment:}~
This research is funded by Vietnam
National Foundation for Science and
Technology Development (NAFOSTED) under
the grant number $103.01$-$2023.16$.
\section*{Appendix A:         
Check for $UV$-finite of      
one-loop form factors}        
After collecting one-loop form factors,
we are going to verify the analytic
expressions by checking for $UV$-finite 
of one-loop form factors. Since there 
is no virtual photon exchanging in 
the loop. The results must be $IR$-finite. 
Moreover, we have no tree-level coupling of 
$A_0$ with $h_0h_0\gamma$. The decay 
channel starts at the contributions of 
one-loop Feynman diagrams.
The results also must be $UV$-finite after 
summing all the contributed diagrams. 
In the course of dimensional regularization,
one-loop Feynman integrals are regularised 
in the space-time dimension $d=4-2\epsilon$
(for $\epsilon \rightarrow 0$ at final result).
It is well-known that each one-loop diagram
contains $UV$-divergent parts which are 
factorised
as $C_{UV}=1/\epsilon -\log(4\pi)
+\gamma_E$ with Euler-Mascheroni constant 
$\gamma_E=0.5772156\cdots$
(see Ref.~\cite{Denner:1991kt} for more detail). 
The sum all diagrams for 
this process is being $UV$-finite. 
For the checks, we take the couplings of the THDM 
Type I as example. Other input parameters are 
listed as follows: $k_{12} = 500^2$ GeV$^2$,
$k_{13} = 300^2$ GeV$^2$, $M_{A_0}^2 = 800^2$ GeV$^2$,
$M_{H_0}^2 = 200^2$ GeV$^2$, 
$M_{H^\pm}^2 = 250^2$ GeV$^2$, 
$M^2 = 100^2$ GeV$^2$,
$t_\beta = 5$, and $s_{\beta - \alpha} = 0.95$.
In the Table~\ref{UV}, we vary 
$C_{UV}$ from $0$ to $10^6$, one finds that 
the results are good stabilities over $14$
digits in agreement. It shows that the one-loop
form factors are $UV$-finite.
\begin{table}[H]
\begin{center}
\centering
\begin{tabular}{|l|l|}  
\hline \hline 
$C_{UV}$
& 
$F (M_{A_0}^2, M_{h_0}^2; 
k_{12}, k_{13}, \cdots)$
\\ \hline
$0$
& 
$
3.1741294086792364
+
1.8029001762071715 \, i
$
\\ \hline
$10^4$
& 
$
3.1741294086792358
+
1.8029001762071712 \, i
$
\\ \hline   
$10^6$
& 
$
3.1741294086792857
+
1.8029001762071708 \, i
$
\\ \hline   
\hline      
\end{tabular}
\caption{
\label{UV}
In this Table, 
checks for the ultraviolet 
finiteness of one-loop form factors. }
\end{center}
\end{table}

\section*{Appendix B:         
The Ward identity checks}     
For the decay process, we have a on-shell 
final photon. Subsequently,
one-loop amplitude follows
the Ward identity as explained in the 
section $3$. In order to confirm 
the identity, two one-loop form 
factors $F_1$ and $F_2$
are collected independently. 
Their relation shown 
in Eq.~\eqref{wardIDD} is then verified
numerically. In the Table~\ref{WardID},
the first column, we change the values of 
$(k_{12}, k_{13})$. While the second and 
the last columns show the numerical 
values for two form factors $F_1$ and $F_2$, 
respectively. For the tests, 
we use the input 
parameters as 
$M_{A_0}^2 = 800^2$ GeV$^2$, 
$M_{H_0}^2 = 200^2$ GeV$^2$, 
$M_{H^\pm}^2 = 250^2$ GeV$^2$, 
$M^2 = 100^2$ GeV$^2$, $t_\beta = 5$, 
and $s_{\beta - \alpha} = 0.95$.
The results are shown in this Table is 
corresponding to the THDM type I  
taken as a typical example.
The results are good stabilities
over $12$ digits. It confirms the 
identity numerically. 
\begin{table}[H]
\begin{center}
\centering
\begin{tabular}{|l|l|l|}  
\hline \hline 
$(k_{12}, k_{13})$ [GeV$^2$]
& 
$(k_1 \cdot k_3) \times F_1$
& 
$- (k_2 \cdot k_3) \times F_2$
\\ \hline
$(-100^2, -200^2)$
& 
$
+
1.6380937118162464
$
& 
$
+
1.6380937118162467
$
\\
$$
& 
$
+
3.315236367886336 \, i
$
& 
$
+
3.315236367886336 \, i
$
\\ \hline
$(+100^2, -200^2)$
& 
$
+
3.401930366245134
$
& 
$
+
3.401930366245136
$
\\
$$
& 
$
+
3.3400883185466217 \, i
$
& 
$
+
3.3400883185466217 \, i
$
\\ \hline   
$(-100^2, +200^2)$
& 
$
+
1.3135551033733244
$
& 
$
+
1.3135551033733248
$
\\
$$
& 
$
-
0.9792390439397629 \, i
$
& 
$
-
0.9792390439397629 \, i
$
\\ \hline   
$(+100^2, +200^2)$
& 
$
+
0.4278406932198653
$
& 
$
+
0.4278406932198655
$
\\
$$
& 
$
-
0.9572229486257442 \, i
$
& 
$
-
0.9572229486257442 \, i
$
\\ 
\hline   
\hline      
\end{tabular}
\caption{
\label{WardID}
In this Table, the numerical checks 
for the Ward identity of one-loop 
amplitude are performed with 
varying $(k_{12}, k_{13})$. }
\end{center}
\end{table}

\end{document}